\documentclass[letterpaper,titlepage,11pt]{article}

\usepackage{amssymb,amsmath,amsfonts}
\usepackage{epsfig}

\def\clock{{\count0=\time
           \divide\count0 60
           \ifnum\count0<10 0\fi\the\count0
           \multiply\count0 -60 \advance\count0 \time
           :\ifnum\count0<10 0\fi \the\count0
         }}
\newcommand{\timestamp}{{\small\vbox{\hbox{\tt\jobname.tex}
\hbox{\the\day/\the\month/\the\year, \clock}}}}

\usepackage{amssymb,amsmath,amsfonts}
\usepackage{epsfig}

\newcommand{\spa}{\ , \ \ }
\newcommand{\R}{\mathbb{R}}
\newcommand{\T}{\mathbb{T}}
\newcommand{\CM}{\mathcal{M}}
\newcommand{\CT}{\mathcal{T}}
\newcommand{\CO}{\mathcal{O}}
\newcommand{\mt}{\mathfrak{t}}
\newcommand{\ms}{\mathfrak{s}}
\newcommand{\nn}{\nonumber}
\newcommand{\bbe}[1]{\mbox{${\mathbb E}^{#1}$}}

\newcommand{\gsim}{\mathrel{\raisebox{-.6ex}{$\stackrel{\textstyle>}{\sim}$}}}
\def\ie{$i.e.$ }
\def\eg{$e.g.$ }
\newcommand{\runinhead}[1]{\vskip .2cm \noindent {\bf #1}}

\long\def\symbolfootnote[#1]#2{\begingroup%
\def\thefootnote{\fnsymbol{footnote}}\footnote[#1]{#2}\endgroup}

\begin{document}

\begin{titlepage}
\begin{flushright}
February 2008
\end{flushright}
\vskip 2cm
\begin{center}
{\bf\Large{ Black Holes in Higher-Dimensional Gravity%
\symbolfootnote[1]{Based on lectures given at the 4th Aegean Summer School on Black Holes,
Sept. 17-22 (2007) Mytiline, Greece.
To be published as part of Lecture Notes in Physics by Springer.}
}} \vskip 1.5cm
{\bf
Niels A. Obers
}
\vskip .5cm
\textit{The Niels Bohr Institute \\
Blegdamsvej 17,  2100 Copenhagen \O,  Denmark}\\
\vskip .5cm
{\tt obers@nbi.dk}
\end{center}
\vskip 1cm

\baselineskip 16pt
%

\begin{center}
{\bf Abstract}
\end{center}
\vskip 0.1cm \noindent
These lectures review some of the recent progress in uncovering the
phase structure of black hole
solutions in higher-dimensional vacuum Einstein gravity. The two classes
on which we focus are Kaluza-Klein black holes, \ie  static solutions with an event
horizon in asymptotically flat spaces with compact directions,
and stationary solutions with an event horizon in asymptotically flat space.
Highlights include the recently constructed multi-black hole configurations
on the cylinder and thin rotating black rings in dimensions higher than five.
The phase diagram that is emerging for each of the two classes will be discussed,
including an intriguing connection that relates the
phase structure of Kaluza-Klein black holes with that of asymptotically flat
rotating black holes.

\end{titlepage} \vfill\eject

\setcounter{equation}{0}

\tableofcontents


\section{Introduction and motivation \label{obesec:intr}}

The study of the phase structure of black objects in higher-dimensional gravity
(see \eg the reviews
\cite{obeKol:2004ww,obeEmparan:2006mm,obeHarmark:2007md,obeEmparan:2008eg})
is interesting for a wide variety of reasons.
First of all, it is of intrinsic interest in gravity where the
spacetime dimension can be viewed as a tunable parameter. In this way one may
discover which properties of black holes are universal and which ones
show a dependence on the dimension. We know for example that the laws
of black hole mechanics are of the former type, while, as will be illustrated
in this lecture, properties such as uniqueness and horizon topology are
of the latter type. In particular, recent research has revealed that as
the dimension increases the phase structure becomes increasingly intricate
and diverse. In this context, another interesting phenomenon that has been
observed is the existence of critical dimensions, above which certain
properties of black holes can change drastically.
Uncovering the phases of black holes is also relevant for the issue
of classical stability of black hole solutions as well as
gravitational phase transitions between different solutions, such as those that involve a
change of topology of the event horizon. Furthermore,
information about the full structure of the static or stationary phases of the theory
can provide important clues about the time-dependent trajectories that interpolate between
different phases.

Going beyond pure Einstein gravity, there are also important motivations originating
from String Theory. String/M-Theory at low energies is described by
higher-dimensional theories of gravity, namely various types of
supergravities. As a consequence, black objects in pure gravity are
often intimately related to black hole/brane solutions in string theory.
These charged cousins and their near-extremal limits
play an important role in the microscopic understanding of black hole entropy
\cite{obeStrominger:1996sh} and other physical properties  (see also \eg the reviews
\cite{obeMathur:2005zp,obeMathur:2005ai}) .
A related application is in the gauge/gravity correspondence
\cite{obeMaldacena:1997re,obeAharony:1999ti},
where the near-extremal limits of these black branes give rise to phases
in the corresponding dual non-gravitational theories at finite temperature.
 In this way, finding new black objects
can lead to the prediction of new phases in these thermal non-gravitational theories
(see \eg \cite{obeAharony:2004ig,obeHarmark:2004ws}).
Finally, if large extra dimensions \cite{obeArkaniHamed:1998rs,obeAntoniadis:1998ig} are
realized in Nature, higher-dimensional black holes would be important as
possible objects to be produced in accelerators or observed in the Universe
(see \eg the review \cite{obeKanti:2004nr}).

In the past seven years, the two classes that have been studied most intensely
are:
\begin{itemize}
\item stationary solutions with an event horizon in asymptotically flat space
\item static solutions with an event horizon in asymptotically flat spaces with compact directions
\end{itemize}
For brevity, we will often refer in this lecture to the first type as
{\it rotating black holes}  and the second
type as {\it Kaluza-Klein black holes}. In this nomenclature, the term
``black hole'' stands for any object with an event horizon, regardless its
horizon topology (\ie not necessarily spherical). We also allow for the
possibility of multiple disconnected event horizons, to which we refer
as {\it multi-black hole solutions}.

For rotating black holes most progress in recent years has been in five
dimensions. Here, it has been found that in addition to the Myers-Perry
(MP) black holes \cite{obeMyers:1986un}, there exist rotating black
rings \cite{obeEmparan:2001wn,obeEmparan:2006mm} and multi-black hole
solutions like black-Saturns and multi-black rings
\cite{obeElvang:2007rd,obeElvang:2007hg,obeIguchi:2007is,obeEvslin:2007fv}
including those with two independent angular momenta
\cite{obePomeransky:2006bd,obeIzumi:2007qx,obeElvang:2007hs}.
All of these are exact solutions which have been obtained with the aid
of special ans\"atze \cite{obeEmparan:2001wk,obeHarmark:2004rm} based on symmetries
and inverse-scattering techniques
\cite{obeBelinsky:1971nt,obeBelinsky:1979,obeBelinski:2001ph,obePomeransky:2005sj}.
We refer in particular to the review \cite{obeEmparan:2006mm} for further details
on the black ring in five dimensions and Ref.~\cite{obeElvang:2007hg}
for a discussion
of the phase diagram in five dimensions for the case of rotating black holes
with a single angular momentum. Moreover, the very recent review
\cite{obeEmparan:2008eg} provides a pedagogical overview of black holes in
higher dimensions, including the more general phase structure of
five-dimensional stationary black holes and solution generating techniques.

Only recently has there been significant progress in exploring the
phase structure of stationary solutions in six and more dimensions
\cite{obeEmparan:2007wm}.
This includes the explicit construction of thin black rings
in six and higher dimensions \cite{obeEmparan:2007wm} based on a perturbative technique
known as matched asymptotic
expansion \cite{obeHarmark:2003yz,obeGorbonos:2004uc,obeKarasik:2004ds,obeGorbonos:2005px,obeDias:2007hg}.
Furthermore, in Ref.~\cite{obeEmparan:2007wm}  the correspondence between
ultra-spinning black holes \cite{obeEmparan:2003sy} and black membranes on a two-torus
was exploited, to take steps towards qualitatively completing the phase diagram of rotating
blackfolds with a single angular momentum. That has led to the proposal
that there is a connection between MP black holes and black rings, and between MP black
holes and black Saturns, through merger transitions involving two kinds
of `pinched' black holes. More generally, this analogy suggests an
infinite number of pinched black holes of spherical topology
leading to a complicated pattern of connections and mergers between phases.
The proposed phase diagram was obtained by importing the present knowledge
of phase of Kaluza-Klein black holes on a two-torus.

For Kaluza-Klein (KK) black holes, most progress has been for the simplest
KK space, namely Minkowski space times a circle. The simplest static solution
of Einstein gravity (in five or more dimensions) in this case is the uniform
black string, which  has a factorized form consisting of
a Schwarzschild-Tangherlini black hole and an extra flat (compactified) direction.
But there are many more phases of KK black holes, which in recent years have been
uncovered by a combination of perturbative techniques (matched asymptotic expansion),
numerical methods and exact solutions. These phases include
non-uniform black strings
(see \cite{obeGubser:2001ac,obeWiseman:2002zc,obeSorkin:2004qq,obeKleihaus:2006ee,obeSorkin:2006wp,obeKleihaus:2007cf}
for numerical results), localized black holes (see
\cite{obeHarmark:2002tr,obeHarmark:2003yz,obeGorbonos:2004uc,obeGorbonos:2005px,obeKarasik:2004ds,obeChu:2006ce,%
obeDias:2007hg,obeKol:2007rx} for analytical results and
\cite{obeSorkin:2003ka,obeKudoh:2003ki,obeKudoh:2004hs} for numerical solutions)
and bubble-black hole sequences \cite{obeElvang:2004iz}. Here recent progress \cite{obeDias:2007hg}
includes the construction of small mass multi-black hole configurations
localized on the circle which in some sense parallel the multi-black hole
configurations obtained for rotating black holes.

All of these static, uncharged phases can be depicted in a two-dimensional phase diagram
\cite{obeHarmark:2003dg,obeKol:2003if,obeHarmark:2003eg} parameterized by the
mass and tension. Mapping out this phase structure has  consequences for the
endpoint of the Gregory-Laflamme instability \cite{obeGregory:1993vy,obeGregory:1994bj}
of the neutral black string, which is a long wavelength instability
 that involves perturbations with an oscillating profile along the direction
of the string. The non-uniform black string phase emerges from the uniform black
string phase at the Gregory-Laflamme point, which is determined by the  (time-independent)
threshold mode where the instability sets in. An interesting property that has
been found in this context is the existence of a critical dimension
\cite{obeSorkin:2004qq} where the transition of the uniform black string into the
 non-uniform black string changes from first order into second order.
Moreover, it has been shown \cite{obeKol:2002xz,obeWiseman:2002ti,obeKol:2003ja,obeSorkin:2006wp}
that the localized black hole phase meets the non-uniform black string phase
in a horizon-topology changing merger point.
Turning to more recent developments, we note that the new multi-black hole
configurations of Ref.~\cite{obeDias:2007hg} raise the question of existence of new non-uniform
black strings. Furthermore, analysis of the three-black hole configuration ~\cite{obeDias:2007hg}
suggests the possibility of a new class of static lumpy black holes in Kaluza-Klein space.

Many of the insights obtained in this simplest case are expected to
carry over as we go to Kaluza-Klein spaces with higher-dimensional compact
spaces \cite{obeKol:2004pn,obeKol:2006vu,obeHarmark:2007md},
although the degree of complexity in these cases will increase substantially.

In summary, recent research has shown that in going from four to higher dimensions
in vacuum Einstein gravity a very rich phase structure
of black holes is observed with fascinating new properties, such as symmetry breaking,
new horizon topologies, merger points and in some cases infinite non-uniqueness.
Obviously one of the reasons for this richer phase structure is that as the dimension increases there are many more degrees of freedom
for the metric. Furthermore, for stationary solutions every time the dimension increases
two units, there is one more orthogonal rotation plane available. Another reason is
the existence of extended  objects in higher dimensions, such as black $p$-branes
(including the uniform black string for $p=1$). Finally, allowing for compact directions
introduces extra scales, and hence more dimensionless parameters in the problem.

The reasons that make the phase structure so rich, such as the increase of the degrees
of freedom and the appearance of fewer symmetries, are those that also
make it hard to uncover. As the overview above illustrates, there has been remarkable
progress in recent years, but we have probably only seen a glimpse of the full
phase structure of black holes in higher-dimensional gravity.
However, the cases considered so far will undoubtedly provide essential clues towards a
more complete picture and will form the basis for further developments into this
fascinating subject.

The outline of these lectures is as follows. To set the stage,
we first give in Sec.~\ref{obesec:uniq} a brief introduction to known uniqueness theorems
for black holes in pure gravity and some prominent cases of non-uniqueness in higher
dimensions.  We also give a short overview of some of the most
important techniques that have been used to obtain
black hole solutions beyond four dimensions.
Then we review the current status for Kaluza-Klein black holes in Secs.~\ref{obesec:kkbh}
and \ref{obesec:mubh}. In particular, Sec.~\ref{obesec:kkbh} presents the main results for
black objects on the cylinder with one event horizon as well as results for Kaluza-Klein black
holes on a two-torus, which will be relevant in the sequel. Sec.~\ref{obesec:mubh}   discusses
the recently constructed multi-black hole configurations on the cylinder.
Then the focus will be turned to rotating black holes in Secs.~\ref{obesec:robh} and
\ref{obesec:phas}. The five-dimensional case will be very briefly reviewed,
but most attention will be given to the recent progress for six and higher
dimensions, including the  construction of thin black rings in Sec.~\ref{obesec:robh}.
We then discuss in Sec.~\ref{obesec:phas} the proposed phase structure for
rotating black holes in six and higher dimensions with a single angular momentum.
The resulting picture builds on an interesting connection to the phase structure of
Kaluza-Klein black holes discussed in the first part.
We end with a future outlook for the subject in Sec.~\ref{obesec:outl}.

\section{Uniqueness theorems and going beyond four dimensions \label{obesec:uniq} }

In this section we first review known black hole uniqueness theorems in
Einstein gravity as well as the most prominent cases of non-uniqueness
of black holes in higher dimensions. We also give an overview of some of the most
important techniques that have been used in finding black hole solutions
beyond four dimensions.

\subsection{Black hole (non-)uniqueness}

The purpose of this lecture is to explore
possible black hole solutions of the vacuum Einstein equations $R_{\mu\nu}=0$
in dimensions $D \geq 4$.
In four-dimensional vacuum gravity, a black hole in an
asymptotically flat space-time is uniquely specified by the ADM mass $M$
and angular momentum $J$ measured at infinity
\cite{obeIsrael:1967wq,obeCarter:1971,obeHawking:1972vc,obeRobinson:1975}.
In particular, in the static case the unique solution is the four-dimensional
Schwarzschild black hole
solution, and for the stationary case it is the Kerr black hole
\begin{subequations}
\label{obeKerr}
\begin{eqnarray}
ds^2 & = & - dt^2 + \frac{\mu r}{\Sigma} (dt + a \sin^2 \theta d \phi)^2 +
\frac{\Sigma}{\Delta} dr^2 + \Sigma d \theta^2 + (r^2+a^2) \sin^2 \theta d \phi^2  , \\
& & \Sigma = r^2 + a^2 \cos^2 \theta \spa \Delta = r^2 - \mu r + a^2 \spa \mu = 2 G
M \spa a = \frac{J}{M} \ .
\end{eqnarray}
\end{subequations}
For $J=0$ this clearly reduces to the Schwarschild black hole, and the angular momentum
is bounded by a critical value $J \leq GM^2 $ (the Kerr bound) beyond which there appears
a naked singularity. The bound is saturated for the extremal Kerr solution  which is
non-singular. The uniqueness in four dimensions fits nicely
with the fact that black holes in four dimensions are known to be
classically stable \cite{obeRegge:1957td,obeZerilli:1971wd,obeTeukolsky:1973ha}
(for further references see also the lecture \cite{obeKodama:2007ph} at this school).

The generalization of the Schwarschild black hole to arbitrary dimension $D$ was
found by Tangherlini \cite{obeTangherlini:1963}, and is given by the metric
\begin{equation}
\label{obeneutbh}
ds^2 = - f dt^2 +  f^{-1} dr^2 + r^2 d \Omega_{D-2}^2 \spa f = 1 - \frac{r_0^{D-3}}{r^{D-3}} \ .
\end{equation}
Here $d\Omega_{D-2}^2$ is the metric element of a $(D-2)$-dimensional
unit sphere with volume $\Omega_{D-2} = 2 \pi^{(D-1)/2}/\Gamma [ (D-1)/2]$.
Since the Newtonian potential $\Phi$ in the weak-field regime $r \rightarrow \infty$
can be obtained from  $g_{tt} = - 1 - 2 \Phi$, this shows that
$\Phi = - r_0^{D-3}/(2 r^{D-3})$. The mass of the black hole is then easily obtained as
\begin{equation}
M = \frac{\Omega_{D-2} (D-2)}{16 \pi G} r_0^{D-3} \ ,
\end{equation}
by using  $ \nabla^2 \Phi =  8 \pi G  \frac{D-3}{D-2} T_{tt} $
and  $ M = \int dx^{D-1} T_{tt} $ where $T_{tt} $ is the energy density.
Uniqueness theorems \cite{obeGibbons:2002bh,obeGibbons:2002av} for
$D$-dimensional ($D > 4$) asymptotically flat space-times state that
the Schwarzschild-Tangherlini black hole solution
 is the only static black hole in pure
gravity. The classical stability of these higher-dimensional black hole solutions
was addressed in Refs.~\cite{obeKodama:2003jz,obeIshibashi:2003ap,obeKodama:2003kk}.

The generalization of the Kerr black hole \eqref{obeKerr} to arbitrary dimension $D$ was
found by Myers and Perry \cite{obeMyers:1986un}, who obtained the metric of
a rotating black hole with angular momenta in an arbitrary number of orthogonal
planes. The Myers-Perry (MP) black hole
is thus specified by the mass and angular momenta
$J_k$ where  $k=1 \ldots r$ with $r = {\rm rank} (SO(D-2))$.
For MP black holes with a single angular momentum,
there is again a Kerr bound $J^2 <
32 G M^3/(27\pi)$ in the five-dimensional case, but for six and more dimensions
the angular momentum is unbounded, and the black hole can be ultra-spinning.
This fact will be important in Secs.~\ref{obesec:robh} and \ref{obesec:phas}.
When there are more than one angular momenta one needs at least one or two
zero angular momenta to have an ultra-spinning regime depending on whether
the dimensions is even or odd \cite{obeEmparan:2003sy}.

Despite the absence of a Kerr bound in six and higher dimensions,
it was argued in \cite{obeEmparan:2003sy} that in six or
higher dimensions the Myers-Perry black hole becomes unstable above
some critical angular momentum thus recovering a dynamical Kerr
bound. The instability was identified as a Gregory-Laflamme
instability by showing that in a large angular momentum limit the
black hole geometry becomes that of an unstable black membrane. This
result is also an indication of the existence of new rotating black
holes with spherical topology, where the horizon is distorted by
ripples along the polar direction. This will be discussed in more detail in
Sec.~\ref{obesec:phas}.
Finally, we note that all of the black hole solutions discussed so far in this
section have an event horizon of spherical topology $S^{D-2}$.

Contrary to the static case, there are no uniqueness theorems
for non-static black holes in pure gravity with $D > 4$.%
\footnote{See \cite{obeHollands:2006rj} for recent progress in this
direction.} On the contrary, there are known cases of
non-uniqueness. The first example of this was found by Emparan and Reall
\cite{obeEmparan:2001wn} and occurs in five dimensions for
stationary solutions in asymptotically flat space-time: for a
certain range of mass and angular momentum there exist both a
rotating MP black hole with $S^3$ horizon \cite{obeMyers:1986un} and
rotating black rings with $S^2 \times S^1$ horizons
\cite{obeEmparan:2001wn}.

As mentioned in the introduction, following the discovery of the
rotating black ring \cite{obeEmparan:2001wn}, further generalizations of these
to black Saturns and multi-black rings have been found in five dimensions.
It is possible that essentially all five-dimensional black holes
(up to iterations of multi-black rings) with two axial Killing vectors have been found
by now%
\footnote{See \cite{obeMorisawa:2004tc,obeHollands:2007aj} for work on how
to determine uniquely the black hole solutions with two symmetry axes.},
but the study of non-uniqueness for rotating black holes
in six and higher dimensions has only recently begun
(see Secs.~\ref{obesec:robh} and \ref{obesec:phas}).

Another case where non-uniqueness has been observed is for Kaluza-Klein
black holes, in particular for black hole solutions that asymptote to
Minkowski space $\CM^{D-1} $ times a circle $S^1$.
Here, the simplest solution one can construct is the uniform black string
which is the $(D-1)$-dimensional Schwarzschild-Tangherlini black hole \eqref{obeneutbh}
plus a flat direction, which has horizon topology $S^{D-3} \times S^1$.
However, at least for a certain range of masses, there
are also non-uniform black strings and black holes that are localized
on the circle, both of which are non-translationally invariant along the circle
direction. All of these solutions, which have in common that they posses
an $SO(D-2)$ symmetry, will be further discussed in Sec.~\ref{obesec:kkbh}.
If one allows for disconnected horizons, then also multi-black hole configurations
localized on the circle are possible, giving rise to a infinite non-uniqueness.
 These will be discussed in Sec.~\ref{obesec:mubh}.
 In addition there are more exotic black hole solutions, called
bubble-black hole sequences \cite{obeElvang:2004iz}, but for simplicity these will not be
further dealt with in this lecture.

More generally, for black hole solutions that asymptote to
Minkowski space $\CM^{D-p} $ times a torus $\T^p$, the simplest class
of solutions with an event horizon are black $p$-branes. The metric
is that of a $(D-p)$-dimensional Schwarzschild-Tangherlini black hole \eqref{obeneutbh}
plus $p$ flat directions. Beyond that there will exist many more phases, which
have only been partially explored.
As an example, we discuss in Sec.~\ref{obesec:torp} the phases of KK black holes
on $\T^2$ that follow by adding a flat direction to the phases of KK black holes
on $S^1$. These turn out to be intimately related to the phase structure of
rotating black holes for $D \geq 6$, as we will see in Sec.~\ref{obesec:phas}.

\subsection{Overview of solution methods \label{obesec:solm}}

We briefly describe here the available methods that have been employed
in order to find the new solutions that are the topic of this lecture.
The main techniques for finding new solutions are as follows.

\runinhead{Symmetries and ans\"atze.} It is often advantageous to
use  symmetries and other physical input to constrain the
form of the metric for the putative solution. In this way one may able to find
an ansatz for the metric that enables to solve the vacuum Einstein equations exactly.
This often involves also a clever choice of coordinate system,
adapted to the symmetries of the problem. This ingredient is also important in
cases where the Einstein equations can only be solved perturbatively around a known
solution (see below).

As an example we note the generalized Weyl ansatz \cite{obeEmparan:2001wk,obeHarmark:2004rm}
for static and stationary solutions with $D-2$ commuting Killing vectors,
in which the Einstein equations simplify considerably. For the static case, this
ansatz is for example relevant for bubble-black hole sequences \cite{obeElvang:2004iz}
in five and six-dimensional KK space. For the stationary case,
it is relevant for rotating black ring solutions
in five dimensional asymptotically flat space.
Another example relevant for black holes and strings on cylinders
is the $SO(D-2)$-symmetric ansatz of
\cite{obeHarmark:2002tr,obeWiseman:2002ti,obeHarmark:2003eg} based on coordinates
that interpolate between spherical and cylindrical
coordinates \cite{obeHarmark:2002tr}.
This has been used to obtain the metric of small black holes on the cylinder
\cite{obeHarmark:2003yz,obeDias:2007hg}.

\runinhead{Solution generating techniques.}
 Given an exact solution there are cases where one can use
solution generating techniques, such as the inverse scattering method, to
generate other new solutions.
See for example Refs.~\cite{obeBelinsky:1971nt,obeBelinsky:1979,obeBelinski:2001ph,obePomeransky:2005sj}
where this method was first used for stationary black hole solutions in five dimensions,
and \cite{obeGiusto:2007fx} for a further solution generating mechanism.

\runinhead{Matched asymptotic expansion.}
In some cases one knows the exact form of the solution in some corner of the moduli space.
Then one may attempt to find the solution in a perturbative expansion around this
(limiting) known solution. This method, called matched asymptotic
expansion \cite{obeHarmark:2003yz,obeGorbonos:2004uc,obeKarasik:2004ds,obeGorbonos:2005px,obeDias:2007hg,obeEmparan:2007wm},
has been very successful. It applies to problems that contain two
(or more) widely separated scales. In particular for black holes, this means
that one solves Einstein equations perturbatively in two different zones,
the asymptotic zone and the near-horizon zone and one thereafter matches
the solution in the overlap region. One example
is that of small black holes on a circle, where the horizon radius of the black
holes is much smaller than the size of the circle (see in particular Sec.~\ref{obesec:mubh}).
Another example is that of thin black rings, where the thickness of the ring is much
smaller than the radius of the ring (see Sec.~\ref{obesec:robh}).

\runinhead{Numerical techniques.}
Since in many cases the Einstein equations
become too complicated to be amenable to analytical methods, even after using
symmetries and ans\"atze, the only way to proceed in the non-linear regime
is to try to solve them numerically. Especially for KK black holes
these techniques have been successfully applied for non-uniform black strings
\cite{obeGubser:2001ac,obeWiseman:2002zc,obeSorkin:2004qq,obeKleihaus:2006ee,obeSorkin:2006wp,obeKleihaus:2007cf}
and localized black holes \cite{obeSorkin:2003ka,obeKudoh:2003ki,obeKudoh:2004hs}
(see Sec.~\ref{obesec:kkbh}).

\runinhead{Classical effective field theory.}
There exists also a classical effective field theory approach
for extended objects in gravity \cite{obeGoldberger:2004jt}. This can be used as a
 systematic low-energy (long-distance) effective expansion which gives results
 only in the region away from the black hole and so it does not provide the corrections
 to the metric near the horizon, but enables one to compute perturbatively
 corrected asymptotic quantities.
This has been successfully applied in \cite{obeChu:2006ce} to obtain the second-order correction
to the thermodynamics of small black holes on a circle. Recently, it was shown
\cite{obeKol:2007rx} that this method is equivalent to matched asymptotic expansion where
the near-horizon zone is replaced by an effective theory. Ref.~\cite{obeKol:2007rx}
also contains an interesting new application of the method to the corrected thermodynamics
of small MP black holes on a circle.

\section{Kaluza-Klein black holes \label{obesec:kkbh}}

 In this section we give a general description of the phases
of Kaluza-Klein (KK) black holes
(see also the reviews \cite{obeHarmark:2007md,obeHarmark:2005pp}).
A $(d+1)$-dimensional Kaluza-Klein black hole will be defined here as a pure gravity
solution with at least one event horizon that asymptotes to
$d$-dimensional Minkowski space times a circle ($\CM^d \times S^1$)
at infinity. We will discuss only static and neutral solutions,
\ie solutions without charges and angular momenta. Obviously, the
uniform black string is an example of a Kaluza-Klein black hole, but
many more phases are known to exist. In particular, we discuss here the non-uniform
black string and the localized black hole phase.
Finally, in anticipation of the connection with the phase structure of
rotating black holes (discussed in Sec.~\ref{obesec:phas}) we also discuss part of the
phases of KK black holes on Minkowski space times a torus ($\CM^{D-2} \times \T^2$).

\subsection{Setup and physical quantities \label{obesec:prel} }

For any space-time which asymptotes to $\CM^d \times S^1$ we can
define the mass $M$ and the tension $\CT$.
These two asymptotic quantities can be
used to parameterize the various phases of Kaluza-Klein black holes
in a $(\mu,n)$ phase diagram, as we review below.

The Kaluza-Klein space  $\CM^d \times S^1$ consists of the time $t$ and
a spatial part which is the cylinder $\R^{d-1}\times S^1$.  The coordinates of
$\R^{d-1}$ are $x^1,...,x^{d-1}$ and the radius $r =\sqrt{\sum_i (x^i)^2 }$.
The coordinate of the  $S^1$ is denoted by $z$ and its circumference is $L$.
It is well known that for static and neutral mass distributions in flat space
$\R^d$ the leading correction to the metric at infinity is given by the mass.
For a cylinder $\R^{d-1}\times S^1$ we instead need two independent asymptotic
quantities to characterize the leading correction to the metric at infinity.

\runinhead{Mass and tension.}
Consider a static and neutral distribution of matter which is localized on a
cylinder $\R^{d-1} \times S^1$. Assume a diagonal energy momentum tensor
with components $T_{tt}$, $T_{zz}$ and $T_{ii}$. Here $T_{tt}$ depends on
$(x^i,z)$ while $T_{zz}$ depends only on $x^i$ because of momentum conservation.
We can then write the mass and tension as
\begin{equation}
M = \int {\rm d} x^d T_{tt} \spa \CT = - \frac{1}{L}  \int {\rm d} x^d  T_{zz} \ .
\end{equation}
From these definitions and the method of equivalent sources, one can
obtain expressions for $M$ and $\CT$ in terms of the leading $1/r^{d-3}$ behavior
of the metric components $g_{tt}$ and $g_{zz}$ around flat space
\cite{obeHarmark:2003dg,obeKol:2003if}. See also Refs.~\cite{obeHarmark:2004ch,obeHarmark:2004ws,%
obeMyers:1999ps,obeTraschen:2001pb,obeTownsend:2001rg,obeKastor:2006ti} for more on the gravitational tension
of black holes and branes.

For a neutral Kaluza-Klein black hole with a single connected
horizon, we can find the temperature $T$ and entropy $S$ directly
from the metric. Together with the mass $M$ and  tension
${\cal{T}}$, these quantities obey the Smarr formula
\cite{obeHarmark:2003dg,obeKol:2003if}
\begin{equation}
\label{obeSmarr1}
(d-1) TS = (d-2)M - L  {\cal{T}} \ ,
\end{equation}
and the first law of thermodynamics \cite{obeTownsend:2001rg,obeKol:2003if,obeHarmark:2003eg}
\begin{equation}
\label{obefirstlaw}
\delta  M = T \delta S + {\cal{T}} \delta L \ .
\end{equation}
This equation includes a ``work'' term (analogous to $p \delta V$)
for variations with respect to the size of the circle at infinity.

It is important to note that there are also examples of Kaluza-Klein
black hole solutions with more than one connected event horizon
\cite{obeHarmark:2003eg,obeElvang:2004iz,obeDias:2007hg}. The Smarr formula \eqref{obeSmarr1}
and first law of thermodynamics \eqref{obefirstlaw} generalize also to these cases.

\runinhead{Dimensionless quantities.}
Since for KK black holes we have an intrinsic scale $L$ it is natural
to use it in order to define dimensionless quantities, which we take as
\begin{equation}
\label{obethemu} \mu = \frac{16\pi G}{L^{d-2}} M \spa
\ms = \frac{16 \pi G}{L^{d-1}} S
\spa \mt = L T  \spa  n = \frac{\CT L}{M} \ .
\end{equation}
Here  $\mu$, $\ms$ and $\mt$ are the rescaled mass, entropy and temperature respectively,
and $n$ is the relative tension.
The relative tension satisfies the bound $0 \leq n \leq d-2$ \cite{obeHarmark:2003dg}.
The upper bound  is due to the Strong Energy Condition whereas the
lower bound was found in \cite{obeTraschen:2003jm,obeShiromizu:2003gc}.
The upper bound can also be understood physically in a more direct
way from the fact that we expect gravity to be an attractive force.
For a test particle at infinity it is easy to see that the
gravitational force on the particle is attractive when $n < d-2$ but
repulsive when $n > d-2$.

The program set forth in \cite{obeHarmark:2003dg,obeHarmark:2003eg} is to
plot all phases of Kaluza-Klein black holes in a $(\mu,n)$ diagram.
Note that it follows from the Smarr formula \eqref{obeSmarr1} and the
first law of thermodynamics \eqref{obefirstlaw} that given a curve $n(\mu)$
in the phase diagram, the entire thermodynamics $\ms(\mu)$ of a phase can be obtained
\cite{obeHarmark:2003dg}.
We also note that the $ (\mu,n)$ phase diagram
appears to be divided into two separate regions \cite{obeElvang:2004iz}. Here, the
region $0 \leq n \leq 1/(d-2)$ contains solutions
without Kaluza-Klein bubbles, and the solutions have a local
$SO(d-1)$ symmetry and reside in the ansatz proposed in \cite{obeHarmark:2002tr,obeHarmark:2003fz}
and proven in \cite{obeWiseman:2002ti,obeHarmark:2003eg}. Solutions of this type,
also referred to as black holes and strings on cylinders
will be reviewed in Sec.~\ref{obesec:bhcy}. Because of the
$SO(d-1)$ symmetry there are only two types of event horizon
topologies: $S^{d-1}$ for the black hole on a cylinder branch and
$S^{d-2} \times S^1$ for the black string. The region $1/(d-2) < n \leq d-2$
contains solutions with Kaluza-Klein bubbles.  This part of the phase diagram,
which is much more densely populated with solutions compared to the lower part,
is the subject of \cite{obeElvang:2004iz}.

\runinhead{Alternative dimensionless quantities.}
The typical dimensionless quantities  used for KK black holes in $D$ dimensions,
are those defined in \eqref{obethemu}.
Instead of these, Ref.~\cite{obeEmparan:2007wm} introduced the following new dimensionless
quantities, more suitable for the analogy with rotating black holes
 (see Sec.~\ref{obesec:phas}), by defining
\begin{equation}
\label{obeeladef}
\ell^{D-3} \propto \frac{L^{D-3}}{G M} \spa  a_H^{D-3} \propto
\frac{S^{D-3}}{(G M)^{D-2}} \spa  \mt_H \propto (G M)^{\frac{1}{D-3}} T  \,.
\end{equation}
In particular, the relation to the dimensionless quantities in \eqref{obethemu} is
given by
\begin{equation}
\label{obeelladef}
\ell  = \mu^{-\frac{1}{D-3}} \spa a_H  =  \mu^{-\frac{D-2}{D-3}} \ms \spa
\mt_H = \mu^{\frac{1}{D-3} } \mt  \ .
\end{equation}
In the KK black hole literature, entropy plots are typically given as $\ms (\mu)$. Instead
of these one can also use \eqref{obeelladef} to
consider the area function $a_H (\ell)$, which is obtained as
\begin{equation}
\label{obeafroms}
a_H (\ell) = \ell^{D-2} \ms (\ell^{-D+3})\ .
\end{equation}
We will employ these alternative quantities when we discuss KK black holes on a torus
in Sec.~\ref{obesec:torp}

\subsection{Black holes and strings on cylinders \label{obesec:bhcy}}

We now discuss the main three types of KK black holes that have
$SO(d-1)$ symmetry, to which we commonly refer as black holes and
strings on cylinders.
These are the uniform black string, the non-uniform black string
and the localized black hole. In Sec.~\ref{obesec:mubh} we will discuss in more
detail the recently obtained multi-black hole configurations on the cylinder.

\subsubsection*{Uniform black string and Gregory-Laflamme instability}

The metric for the uniform black string in $D=d+1$ space-time
dimensions is
\begin{equation}
\label{obeublstr} ds^2 = - f dt^2 + f^{-1} dr^2 + r^2 d\Omega_{d-2}^2
+ dz^2 \spa f=1-\frac{r^{d-3}}{r_0^{d-3}} \ ,
\end{equation}
where $d\Omega_{d-2}^2$ is the metric element of a $(d-2)$-dimensional
unit sphere.  The metric \eqref{obeublstr} is found by taking the
$d$-dimensional Schwarzschild-Tangherlini static black hole \eqref{obeneutbh}
solution \cite{obeTangherlini:1963} and adding a flat $z$ direction, which is the direction
parallel to the string. The event horizon is located at $r=r_0$
and has topology $S^{d-2}\times \R$.

\runinhead{Gregory-Laflamme instability.}
Gregory and Laflamme  found in 1993 a long wavelength instability
for black strings in five or more dimensions \cite{obeGregory:1993vy,obeGregory:1994bj}.
The mode responsible for the instability propagates along the direction of
the string, and develops an exponentially growing time-dependent
part when its wavelength becomes sufficiently long. The Gregory-Laflamme mode is a
linear perturbation of the metric \eqref{obeublstr}, that can be written as
\begin{equation}
\label{obepertmet} g_{\mu\nu} + \epsilon h_{\mu\nu} \ .
\end{equation}
Here $g_{\mu\nu}$ stands for the components of the unperturbed
black string metric \eqref{obeublstr},  $\epsilon$ is a small parameter and
$h_{\mu\nu}$ is the metric perturbation
\begin{equation}
\label{obeGLmode1} h_{\mu\nu} = \Re \left\{ \exp \left( \frac{\Omega
t}{r_0} + i \frac{kz}{r_0} \right) P_{\mu\nu} ( r/r_0) \right\} \ ,
\end{equation}
where the symbol $\Re$ denotes the real part.
The statement that the perturbation $h_{\mu\nu}$ of $g_{\mu\nu}$ satisfies the
Einstein equations of motion can be stated as the differential operator equation
\begin{equation}
\label{obelicheq} \Delta_L h_{\mu\nu} = 0 \ ,
\end{equation}
where $(\Delta_L)_{\mu \nu\rho \sigma} = - g_{\mu \rho}g_{\nu\sigma} D_\kappa
D^\kappa + 2 R_{\mu\nu\rho\sigma}$ is the Lichnerowitz operator for the background metric
$g_{\mu\nu}$. The resulting Einstein equations for the GL mode
can be found \eg in the appendix of the review \cite{obeHarmark:2007md}.%
\footnote{Various methods and different gauges have been employed to derive the
differential equations for the GL mode.
See Ref.~\cite{obeKol:2006ga} for a nice summary of these,
including a new derivation (see also \cite{obeKol:2006ux}).}
Solution of these equations \cite{obeGregory:1993vy,obeGregory:1994bj} shows
that there is an unstable mode for any
wavelength larger than the critical wavelength
\begin{equation}
\label{obelambgl}
\lambda_{\rm GL} = \frac{2\pi r_0 }{k_c}
~.
\end{equation}
for which $\Omega=0$ in \eqref{obeGLmode1}.
The values of $k_c$ for $d=4,...,14$, as obtained in
\cite{obeGregory:1993vy,obeGubser:2001ac,obeSorkin:2004qq}, are listed \eg in
Table 1 of \cite{obeHarmark:2007md}. The critical wave-number $k_c$
marks the lower bound of the possible wavelengths for which there is an
unstable mode and is called the threshold mode.
It is a time-independent mode of the form
$ h_{c,\mu\nu} \sim  \exp ( i k_c z/r_0 )$. In particular, this suggests the existence of a
 static non-uniform black string.

\runinhead{GL mode of the compactified uniform black string.}
Since we wish to consider the uniform black string in KK space, we
now discuss what happens to the GL instability when $z$ is
a periodic coordinate with period  $L$. The Gregory-Laflamme mode \eqref{obeGLmode1}
cannot obey the correct periodic boundary condition on $z$
if $L < \lambda_{\rm GL}$, with $\lambda_{\rm GL}$ given by
\eqref{obelambgl}. On the other hand, for $L > \lambda_{\rm GL}$,
we can fit the Gregory-Laflamme mode into the compact direction
with the frequency and wave number $\Omega$ and $k$
in \eqref{obeGLmode1} determined by the ratio $r_0/L$.
Translating this in terms of the mass of the neutral black string,
one finds the critical Gregory-Laflamme mass
\begin{equation}
\label{obemugl} \mu_{\rm GL} = (d-2)\Omega_{d-2} \left(
\frac{k_c}{2\pi} \right)^{d-3}
~.
\end{equation}
For $\mu < \mu_{\rm GL}$ the Gregory-Laflamme mode can be fitted
into the circle, and the compactified neutral uniform black string
is unstable. For $\mu > \mu_{\rm GL}$, on the other hand, the
Gregory-Laflamme mode is absent, and the neutral uniform black
string is stable. For $\mu=\mu_{\rm GL}$ there is a marginal mode
which signals the emergence of a new branch of black string solutions which are
non-uniformly distributed along the circle. See \eg Table 2 in \cite{obeHarmark:2007md}
 for the values of $\mu_{\rm GL}$ for $ 4 \leq d \leq 14$.

The large $d$ behavior of $\mu_{\rm GL}$ was examined numerically in
 \cite{obeSorkin:2004qq} and analytically in \cite{obeKol:2004pn}.
We also note that there is an interesting correspondence between the
Rayleigh-Plateau instability of long fluid cylinders and the Gregory-Laflamme
instability of black strings \cite{obeCardoso:2006ks,obeCardoso:2006sj}.
In particular, the critical wave numbers $k_{RP}$ and
$k_{c}$ agree exactly at large dimension $d$  (scaling both as $\sqrt d$ for
$d \gg 1$).

\subsubsection*{Non-uniform black string}

It was realized in \cite{obeGubser:2001ac} (see also \cite{obeGregory:1988nb})
 that the classical
instability of the uniform black string for $\mu < \mu_{\rm GL}$
implies the existence of a marginal (threshold) mode at $\mu =\mu_{\rm GL}$,
which again suggests the existence of a new branch of solutions.

The new branch, which is called the non-uniform string branch, has
been found numerically in \cite{obeGubser:2001ac,obeWiseman:2002zc,obeSorkin:2004qq}.
This branch of solutions has the same horizon topology  $S^1 \times S^{d-2}$
as the uniform string, which is expected since the non-uniform string
 is continuously connected to the uniform black string. In particular, it emerges
 from the uniform black string in the point $(\mu,n) = (\mu_{\rm GL},1/(d-2))$ and
 has $n < 1/(d-2)$
 Moreover, the solution is
 non-uniformly distributed in the circle-direction $z$ since there
is an explicit dependence in the marginal mode on this direction.

More concretely, considering the non-uniform black string branch for
$|\mu-\mu_{\rm GL}| \ll 1$ one obtains for the relative tension the
behavior
\begin{equation}
\label{obenofmu}
n(\mu) = \frac{1}{d-2} - \gamma ( \mu - \mu_{\rm GL}) + \CO ( (
\mu - \mu_{\rm GL})^2 ) \ .
\end{equation}
Here $\gamma$ is a number representing the slope of the curve
that describes the non-uniform string branch near $\mu=\mu_{\rm GL}$
(see Table 3 in \cite{obeHarmark:2007md} for the values of $\gamma$
for $4 \leq d \leq 14$ obtained from the data
of \cite{obeGregory:1993vy,obeGregory:1994bj,obeGubser:2001ac,obeWiseman:2002zc,obeSorkin:2004qq}).

The qualitative behavior of the non-uniform string branch depends on
the sign of $\gamma$. If $\gamma$ is positive, then the branch
emerges at the mass $\mu=\mu_{\rm GL}$ with increasing $\mu$ and
decreasing $n$. If instead $\gamma$ is negative the branch emerges
at $\mu=\mu_{\rm GL}$ with decreasing $\mu$ and decreasing $n$.
To see what this means for the entropy we note
that from \eqref{obenofmu} and the first law  of thermodynamics one finds
\begin{equation}
\label{obenuentro}
\frac{\ms_{\rm nu} ( \mu )}{\ms_{\rm u}  ( \mu )}
= 1 - \frac{(d-2)^2}{2(d-1)(d-3)^2} \frac{\gamma}{\mu_{\rm GL}}
(\mu-\mu_{\rm GL})^2 + \CO ( (\mu - \mu_{\rm GL})^3 ) \ ,
\end{equation}
where $\ms_{\rm u} ( \mu )$
($\ms_{\rm nu} ( \mu )$) refers to the rescaled entropy of the
uniform (non-uniform) black string branch.
It turns out that $\gamma$ is positive for $d \leq 12$ and
negative for $d \geq 13$ \cite{obeSorkin:2004qq}. Therefore, as
discovered in \cite{obeSorkin:2004qq}, the non-uniform black string
branch has a qualitatively different behavior for small $d$ and
large $d$, \ie the system exhibits a critical dimension  $D=14$.
In particular, for $d \leq 12$ the non-uniform branch near the
GL point has $\mu > \mu_{\rm GL}$ and lower entropy than that
of the uniform phase, while for $d > 13$ it has $\mu < \mu_{\rm GL}$
and higher entropy. It also follows from \eqref{obenuentro} that for all
$d$ the curve $\ms_{\rm nu} (\mu)$ is tangent to the curve $\ms_{\rm u}(\mu)$
at the GL point.

A large set of numerical data for the non-uniform branch, extending into the
strongly non-linear regime, have been obtained
in Refs.~\cite{obeWiseman:2002zc,obeKudoh:2004hs} for six dimensions (\ie $d=5$)
in Ref.~\cite{obeKleihaus:2006ee} for five dimensions (\ie $d=4$) and
for the entire range $d \leq 5 \leq 10$ in
Ref.~\cite{obeSorkin:2006wp}. For $d=5$, these data are displayed
in the $(\mu,n)$ phase diagram \cite{obeHarmark:2003dg} of Fig.~\ref{obefig1}.

\subsubsection*{Localized black holes}

On physical grounds, it is natural to expect a branch of neutral
black holes in the space-time $\CM^d \times S^1$ with
event horizon of topology $S^{d-1}$. This branch is called the localized black hole
branch, because the $ S^{d-1}$ horizon is localized on the $S^1$ of the Kaluza-Klein space.

Neutral black hole solutions in the space-time $\CM^3 \times S^1$
were found and studied in
\cite{obeMyers:1987rx,obeBogojevic:1991hv,obeKorotkin:1994dw,obeFrolov:2003kd}.
However, the study of black holes
in the space-time $\CM^{d} \times S^1$ for $d \geq 4$ is relatively new.
The complexity of the problem stems from the fact that such
black holes are not algebraically special \cite{obeDeSmet:2002fv}
and moreover from the fact that the solution cannot be found using
a Weyl ansatz since the number of Killing vectors is too small.

In \cite{obeHarmark:2003yz,obeGorbonos:2004uc,obeGorbonos:2005px} the metric of
small black holes, \ie black holes with mass $\mu \ll 1$, was
found analytically using the method of matched asymptotic expansion.
The starting point in this construction is the fact that
as $\mu \rightarrow 0$, one has $n \rightarrow 0$ so that
the localized black hole solution becomes more and more like a
$(d+1)$-dimensional Schwarzschild black hole in this limit.
 For $d=4$, the second order correction to the metric and
thermodynamics have been studied in \cite{obeKarasik:2004ds}.
More generally, the second order correction to the thermodynamics was obtained
in Ref.~\cite{obeChu:2006ce} (see also \cite{obeKol:2007rx}) for all $d$ using an
effective field theory formalism in which the structure of the black hole is encoded in the coefficients of
operators in an effective worldline Lagrangian.

The first order result of \cite{obeHarmark:2003yz} and second order result of \cite{obeChu:2006ce}
can be summarized by giving the first and second order corrections
to the relative tension $n$ of the localized black hole branch
as a function of $\mu$%
\begin{equation}
\label{obebhslope}
n = \frac{(d-2)\zeta(d-2)}{2(d-1)\Omega_{d-1}} \mu -
\left( \frac{(d-2)\zeta(d-2)}{2(d-1)\Omega_{d-1}} \mu \right)^2 + \CO (\mu^3) \ ,
\end{equation}
where $\zeta(p) = \sum_{n=1}^\infty n^{-p}$ is the Riemann zeta function.
The corresponding correction to the thermodynamics can be found \eg in Eq.~(3.18)
of \cite{obeHarmark:2007md}.

The black hole branch has been studied numerically
for $d=4$ in \cite{obeSorkin:2003ka,obeKudoh:2004hs}
and for $d=5$ in \cite{obeKudoh:2003ki,obeKudoh:2004hs}.
For small $\mu$, the impressively accurate data of
\cite{obeKudoh:2004hs} are consistent with the analytical results
of \cite{obeHarmark:2003yz,obeGorbonos:2004uc,obeKarasik:2004ds}. The results of
 \cite{obeKudoh:2004hs} for $d=5$ are displayed
in a $(\mu,n)$ phase diagram in Figure \ref{obefig1}.

\subsection{Phase diagram and copied phases \label{obesec:pdia}}

\begin{figure}[ht]
\begin{picture}(0,0)(0,0)
\put(345,30){$\mu$} \put(118,230){$n$}
\end{picture}
\centerline{\epsfig{file=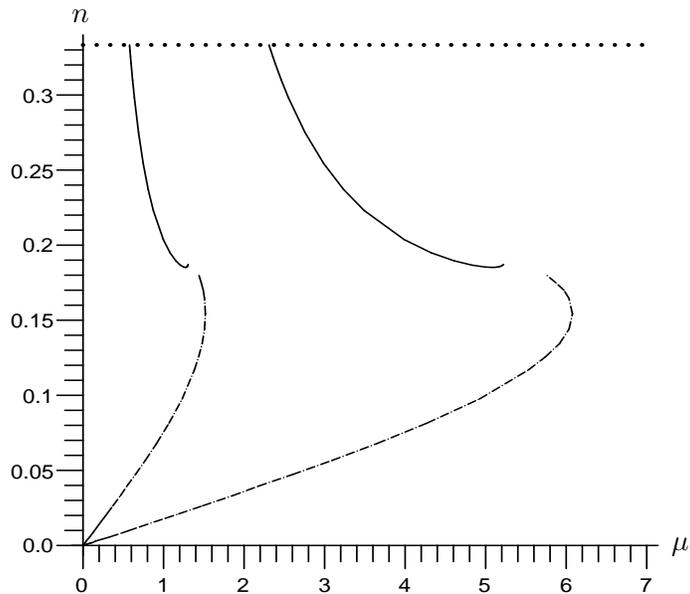,width=9cm,height=8cm} }
\caption{Black hole and string phases for $d=5$, drawn in the $(\mu,n)$ phase diagram.
The horizontal (dotted) line is the uniform string branch.
The rightmost solid branch emanating from this at the Gregory-Laflamme point
is the non-uniform string branch and the  rightmost dashed branch starting in the origin is the
localized black hole branch.
 The solid and dashed
branches to the left are the $k=2$ copies of the non-uniform and localized branch.
 The results strongly suggest that the black hole and non-uniform black string branches meet.}
\label{obefig1}
\end{figure}

In Figure \ref{obefig1} the $(\mu,n)$ diagram for $d=5$ is displayed, which is
one of the cases most information is known.
We have shown the complete non-uniform
branch, as obtained numerically by Wiseman \cite{obeWiseman:2002zc}, which
emanates at $\mu_{\rm GL} = 2.31$ from the uniform branch that has $n=1/3$.
These data were first incorporated into the $(\mu,n)$ diagram
in Ref.~\cite{obeHarmark:2003dg}. For the black hole branch we have plotted the
 numerical data of Kudoh and Wiseman \cite{obeKudoh:2004hs}.
 It is evident from the figure that this branch has an
approximate linear behavior for a fairly large range of $\mu$ close to
the origin and the numerically obtained slope agrees very well with the
analytic result \eqref{obebhslope}.

\runinhead{Merger point.}
The figure strongly suggests that the localized black hole branch meets
with the non-uniform black string branch in a topology changing
transition point, which is the scenario earlier suggested by
Kol \cite{obeKol:2002xz} (see \cite{obeHarmark:2003eg} for a list of scenarios).
For this reason, it seems reasonable to expect that the localized black hole
branch is connected with the non-uniform string branch in any
dimension. This means that we can go from the uniform black string
branch to the localized black hole branch through a connected
series of static classical geometries. The point in which the two
branches are conjectured to meet is called the merger point.

\runinhead{Copied phases.} In \cite{obeHarmark:2003eg} it was shown that one can generate new
solutions by copying solutions on the circle several times,
following an idea of Horowitz \cite{obeHorowitz:2002dc}. This works
for solutions which vary along the circle direction (\ie in the $z$
direction), so it works both for the black hole branch and the
non-uniform string branch. Let $k$ be a positive integer. Then if
we copy a solution $k$ times along the circle we get a new
solution with the following parameters:
\begin{equation}
\label{obecoptrans} \tilde{\mu} = \frac{\mu}{k^{d-3}}\spa \tilde{\ms} =
\frac{\ms}{k^{d-2}} \spa \tilde{\mt} = k \mt   \spa \tilde{n}
= n  \ .
\end{equation}
See Ref.~\cite{obeHarmark:2003eg} for the corresponding expression of
the metric of the copies in the $SO(d-1)$-symmetric ansatz.
Using the transformation \eqref{obecoptrans}, one easily sees that
the non-uniform and localized black hole branches depicted in
Fig.~\ref{obefig1} are copied infinitely many times in the
$(\mu,n)$ phase diagrams, and we have depicted the $k=2$ copy in this figure.

\runinhead{General dimension.} The six-dimensional phase diagram displayed in
Fig.~\ref{obefig1} is believed to be representative for the black string/localized
black hole phases on $\CM^{D-1} \times S^1$ for
all $ 5 \leq D \leq 13$. Here the upper bound follows from fact that, as mentioned above,
there is a critical dimension  $D=13$ above which the behavior of the non-uniform
 black string phase is qualitatively different \cite{obeSorkin:2004qq}.
The phase diagram for $D \geq 14$ is much poorly known in comparison,
since there are no data like fig.~\ref{obefig1} available for the
localized and non-uniform phases, only the asymptotic behaviors. However,
we do know from \eqref{obenofmu} that the non-uniform branch will extend
to the left (lower values of $\mu$) as it emerges from the GL point, and
on general grounds is expected to merge again with the localized black hole branch.

\subsection{KK phases on $\T^2$ from phases on $S^1$ \label{obesec:torp} }

We show here how one can
translate the known results for KK black holes on the circle
(\ie on $\CM^{D-2} \times S^1$) to results for
KK black holes on the torus ($\ie$ on $\CM^{D-2} \times \T^2$).
The resulting phases are relevant in connection with the phases
of rotating black holes in asymptotically flat spacetime,
as shown in Sec~\ref{obesec:phas}.

We recall first the definitions of dimensionless quantities in \eqref{obethemu}.
While these quantities were originally
introduced in \cite{obeHarmark:2003dg,obeHarmark:2004ws} for black holes on
a KK circle of circumference $L$, we may
similarly use these definitions for KK black holes in $D$ dimensions
with a square torus of side lengths $L$, to which we restrict in the following.
Likewise, we can use the alternative dimensionless quantities \eqref{obeelladef} for
that case.

\runinhead{Map from circle to torus compactification.}
We first want to establish a map for these dimensionless quantities from
KK black holes on $\CM^{D-2} \times S^1$ (denoted with hatted quantities) to
those for KK black holes on $\CM^{D-2} \times \T^2$ (denoted with unhatted quantities),
obtained by adding an extra compact direction of size $L$.
Suppose we are given an entropy function $\hat \ms(\hat \mu)$ for a phase of KK
black holes on $\CM^{D-2} \times S^1$.
Any such phase lifts trivially to a phase of KK black holes on
$\CM^{D-2} \times \T^2$ that is uniform in one of the torus directions.
We show below how to obtain the function $a_H(\ell)$ for the latter in terms of
$\hat \ms (\hat \mu)$ of the former. In the following we will use the notation
$D = n +4$.

It is not difficult to see that in terms of the original dimensionless quantities
\eqref{obethemu} we have the simple mapping
\begin{equation}
\label{obemsmap} \mu = \hat \mu \spa \ms = \hat \ms \spa \mt = \hat \mt \ .
\end{equation}
It then follows from \eqref{obeelladef} and \eqref{obeafroms} that the
area function $a_H (\ell)$ of KK black holes on $\CM^{D-2} \times \T^2$ is
obtained via the mapping relation
\begin{equation}
\label{obeaHmap}
a_H (\ell) = \ell^{n+2} \hat \ms (\ell^{-n-1} ) \ .
\end{equation}

\begin{figure}[ht]
\begin{picture}(0,0)(0,0)
\put(345,37){$\ell$} \put(115,260){$a_H$}
\end{picture}
\centerline{\includegraphics[width=9cm]{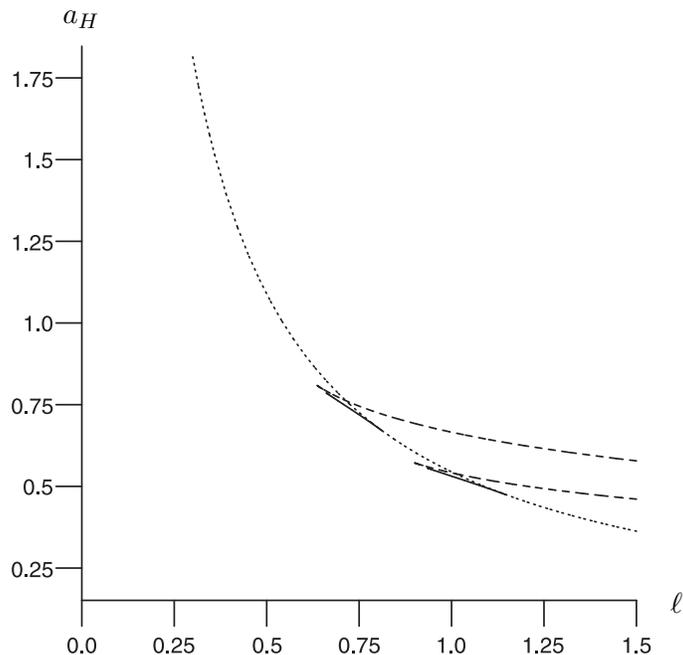}} \caption{
\small $a_H (\ell)$ phase diagram in seven dimensions ($\CM^5 \times
\T^2$) for Kaluza-Klein black hole phases with one uniform
direction. Shown are the uniform black membrane phase (dotted), the
non-uniform black membrane phase (solid) and the localized black
string phase (dashed). For the latter two phases, we have also shown
their $k=2$ copy. The non-uniform black membrane phase emanates from
the uniform black membrane phase at the GL point $\ell_{\rm GL} =
0.811 $, while the $k=2$ copy starts at the 2-copied GL point
$\ell_{\rm GL}^{(2)} = \sqrt{2} \ell_{\rm GL} =1.15 $.
This figure is representative for
the phase diagram of phases on $\CM^{D-2} \times \T^2$ for all $ 6
\leq D \leq 14$. Reprinted from Ref.~\cite{obeEmparan:2007wm}.
 \label{obefig:KKphases7}}
\end{figure}

\runinhead{Application to known phases.}
Using now the entropy function
$\hat \ms_{\rm uni} (\hat \mu) \sim \hat \mu^{ \frac{n}{n-1}}$
of the uniform black string  in  $\CM^{n+2} \times S^1$
we get from \eqref{obeaHmap} the result
\begin{equation}
\label{obeaHmem}
a_H^{\rm ubm} (\ell) \sim \ell^{- \frac{2}{n-1}} \ .
\end{equation}
for the uniform black membrane (ubm) $ 4 +n$ dimensions.
Furthermore, using that for small $\mu$ (or equivalently large $\ell $)
the entropy of the localized black hole in $\CM^{n+2} \times S^1$ is
$
\hat \ms_{\rm loc} (\hat \mu) \sim \hat \mu^{ \frac{n+1}{n}}$
we find via the map \eqref{obeaHmap} the result
\begin{equation}
\label{obeaHstr}
a_H^{\rm lbs} (\ell) \sim \ell^{- \frac{1}{n}}   \qquad (\ell
\rightarrow \infty) \ ,
\end{equation}
for the large $\ell$ limit of the localized black string  (lbs) in $4+n$ dimensions.

Finally, for the non-uniform string
in $\CM^{n+2} \times S^1$ dimensions we use \eqref{obenuentro} to obtain
\begin{equation}
\label{obeahnum}
a_H^{\rm nubm} (\ell) = a_H^{\rm ubm}  (\ell) \left[ 1 - \frac{n^2 (n+1)}{2 (n-1)^2}
\frac{ \gamma_{n+2} }{\ell_{\rm GL}^{n+4}} ( \ell - \ell_{\rm GL})^2
+ \CO ( (\ell - \ell_{\rm GL})^3 ) \right] \ ,
\end{equation}
for the non-uniform black membrane (nubm).
Here, $\ell_{\rm GL} = (\mu_{{\rm GL},n+2})^{-\frac{1}{n+1}}$
is the critical GL wavelength in terms of the dimensionless GL mass
$\mu_{{\rm GL},d}$ given in \eqref{obemugl} and $\gamma_d$ the coefficient in \eqref{obenofmu}.

\runinhead{Copies.} As remarked in Sec.~\ref{obesec:pdia} the
 localized black hole and non-uniform black string phase
on $\CM^{n+2} \times S^1$ have copied phases with multiple non-uniformity or
multiple localized black objects. From the map \eqref{obecoptrans} we then find
using \eqref{obemsmap} and the definitions \eqref{obeelladef} that
 corresponding copied phases of KK black holes on the torus obey
the transformation rule
\begin{equation}
\label{obecopy}
\tilde \ell = k^{\frac{n-1}{n+1}} \ell \spa
\tilde a_H = k^{- \frac{2}{n+1}} a_H \spa \tilde{\mt}_H = k \mt_H \  .
\end{equation}

\runinhead{Seven-dimensional phase diagram.}
As an explicit example, we give the  mapping that can be used to convert the known results
for KK black holes on $\CM^5 \times S^1$ to  KK black holes on $\CM^5 \times \T^2$.
\begin{equation}
(\ell , a_H ) = ( \hat \mu^{-1/4}, \hat \mu^{-5/4} \hat \ms) \ .
\end{equation}
This can be used to convert plots of points $(\hat \mu , \hat \ms)$
(see \eg \cite{obeHarmark:2007md}) for six-dimensional KK black holes on a circle
to the phase diagram in Fig.~\ref{obefig:KKphases7} for seven-dimensional
KK black holes (with one uniform direction) on a torus. It includes the uniform black membrane,
the black membrane with one uniform and one non-uniform direction, and the
black string localized in one of the circles of $\T^2$. The figure also includes
the $k=2$ copies obtained from these data and the map \eqref{obecopy}.
 Both the uniform black membrane phase and the localized black string phase extend to $\ell \rightarrow \infty$
where they obey the behavior \eqref{obeaHmem} and \eqref{obeaHstr}
respectively with $n=3$.

\section{Multi-black hole configurations on the cylinder \label{obesec:mubh} }

We now turn to the construction of multi-black hole configurations on the cylinder,
recently obtained in Ref.~\cite{obeDias:2007hg}.
In Sec.~\ref{obesec:pdia} we already encountered a special subset of these,
namely the copied phases of the localized black hole branch, corresponding to
multi-black hole configurations in which all black holes have the same mass.
Here, we describe the main points of the construction of more general
multi-black hole configurations \cite{obeDias:2007hg} using matched asymptotic expansion. We
also show how
the thermodynamics of these configurations can be understood from a Newtonian point
of view. Finally we comment on the consequences of these configurations for the
phase diagram of KK black holes.

\subsection{Construction of multi-black hole solutions \label{obesec:cons} }

The copies of the single-black hole localized on the circle, correspond
to a multi-black hole configurations of equal mass black holes that are
spread with equal distance from each other on the circle.
Beyond these, there exist more general multi-black hole configurations which have recently been
considered in Ref.~\cite{obeDias:2007hg}. These solutions correspond to having
several localized black holes of different sizes located at different
points along the circle direction of the cylinder $\R^{d-1} \times
S^1$. The location of each black hole is such that the total force
on each of them is zero, ensuring that they are in equilibrium. It
is moreover necessary for being in equilibrium that the black holes
are all located in the same point in the $\R^{d-1}$ part of the
cylinder.

The metric constructed in Ref.~\cite{obeDias:2007hg} are solutions to the Einstein equations to
first order in the mass. More precisely, they are valid in a regime where
the gravitational interaction between any one of the black holes and the others
(and their images on the circle) is small. The solutions in Ref.~\cite{obeDias:2007hg} thus
describe the small mass limit of these multi-black hole configurations
on the cylinder, or equivalently they can be said to describe the situation where the
black holes are far apart. The method used for  solving the
Einstein equations  is the one of matched
asymptotic expansion
\cite{obeHarmark:2003yz,obeGorbonos:2004uc,obeKarasik:2004ds,obeGorbonos:2005px,obeDias:2007hg}.
The particular construction follows the approach of
\cite{obeHarmark:2003yz} where it was used to find the metric of a small black hole
 on the cylinder based on an ansatz for the metric found in \cite{obeHarmark:2002tr}.

\runinhead{General idea and starting point.}
We describe here the general idea behind constructing the new solutions for
multi-black hole configurations on the $d$-dimensional cylinder $\R^{d-1} \times S^1$.
The configuration under consideration is that of $k$ black holes placed
at different locations $z_i^*$, $i=1 \ldots k$  in the same point of the
$\R^{d-1}$ part of the cylinder.
We write $M$ as the total mass of all of the black holes
and define $\nu_i$ as the fraction of mass of the $i^{\rm th}$ black hole, \ie
\begin{equation}
\label{obemu} M_i = \nu_i M \,,\qquad \sum_{i=1}^k \nu_i = 1 \,,
\end{equation}
where $M_i$ is the mass of the $i^{\rm th}$ black hole. Note that $0
< \nu_i \leq 1$.

The matched asymptotic expansion  is suitable when there are two widely separated
scales in the problem. Here they are the size (mass) of each of the black holes,
(all of which are taken of the same order) and the length of the circle direction.
In particular, we assume that all black holes have a horizon radius
(of the same order) which is small compared to the length of the circle.

The constructing of  the solution then proceeds in the following  steps%
\footnote{Here we use the coordinate $R$ which is part of the two-dimensional
coordinate system $(R,v)$ introduced in Ref.~\cite{obeHarmark:2002tr} that
interpolates between cylindrical coordinates $(r,z)$ and spherical coordinates
$(\rho,\theta)$. In terms of $F(r,z)$ in \eqref{obedefF} we have
$R (r,z) \propto F (r,z)^{-1/(d-3)}$. Note that
Refs.~\cite{obeHarmark:2002tr,obeHarmark:2003yz,obeDias:2007hg} set $L= 2\pi$, which we choose
not to do here for pedagogical clarity.}
\begin{itemize}
\item Step 1: Find a metric corresponding to the Newtonian gravitational potential
sourced by a configuration of small black holes on the cylinder.
This metric is valid in the region $R \gg R_0$.
\item Step 2: Consider the Newtonian solution close to the sources, \ie in
the overlap region $R_0 \ll R \ll L $.
\item Step 3: Find a general solution near a given event horizon and match
this solution to the metric in the overlap region found in Step 2.
The resulting solution is valid in the region $R_0 \leq R \ll L$.
\end{itemize}
With all these three steps implemented, we have a complete solution
for all of the spacetime outside the event horizon.
We refer to Ref.~\cite{obeDias:2007hg} for further details, including
the explicit form of the first-order corrected metric and  thermodynamics
of the resulting multi-black hole configurations, but present
some of the easy steps here.

\runinhead{Newtonian potential.}
Following the discussion in Sec.~\ref{obesec:prel}
for static solutions on the cylinder the two relevant components of
the stress tensor are $T_{tt}$ and $T_{zz}$. These components source
the two gravitational potentials \cite{obeHarmark:2003dg}
\begin{equation}
\label{obepots} \nabla^2 \Phi = 8\pi G \frac{d-2}{d-1}
T_{tt} \spa \qquad \nabla^2 B =  \frac{8\pi G}{d-1} T_{zz}  \ ,
\end{equation}
where $G$ is the $(d+1)$-dimensional Newton constant.
In the limit of small total mass, we have that $B/(G M) \rightarrow 0$
for $M \rightarrow 0$ which means \cite{obeHarmark:2003yz,obeDias:2007hg}
that we can neglect the binding energy potential $B$ as compared
to the mass density potential $\Phi$. One thus only needs to consider the
potential $\Phi$, $\ie$  Newtonian gravity.

For the multi-black hole configuration described above, it is not difficult
find the solution for $\Phi$ using the method of images in terms of  $(r,z)$ coordinates of
the cylinder. One finds
\begin{eqnarray}
\Phi (r,z) = - \frac{8 \pi G M  }{(d-1) \Omega_{d-1}}
F (r,z) \,,
 \label{obePhi}
\end{eqnarray}
with
\begin{eqnarray}
F (r,z) = \sum_{i=1}^k \sum_{m=-\infty}^\infty \frac{\nu_i}{ [ r^2 +
(z - z_i^* - L m )^2]^{\frac{d-2}{2}}}  \, ,
 \label{obedefF}
\end{eqnarray}
so that the potential \eqref{obePhi} describes the Newtonian gravitational
potential sourced by the multi-black hole configuration.

One can now study how the potential $\Phi$ looks when going near the sources.
To achieve this it is useful  to define for
the $i^{\rm th}$ black hole the spherical coordinates $\rho$ and
$\theta$ by
\begin{equation}
\label{oberhotheta} r = \rho \sin \theta \spa z-z^*_i = \rho \cos
\theta \, ,
\end{equation}
where $\theta$ is defined in the interval $[0,\pi]$. In terms of these
coordinates one finds that $F(r,z)$ in \eqref{obedefF} can be expanded as
\begin{eqnarray}
F (\rho,\theta) = \nu_i \rho^{-(d-2)} +  \Lambda^{(i)} +
\Lambda_1^{(i)} \cos\theta \,\rho  + {\cal O}\left(\rho^2 \right) \,,
 \label{obelim F}
\end{eqnarray}
for $\rho \ll 1$, where
\begin{eqnarray}
 \Lambda^{(i)} & = &   \frac{1}{L^{d-2}} \Big( \nu_i \,2\zeta (d-2)\nonumber
 \\
  & &  +   \sum_{\substack{j=1 \\j\neq i } }^k
  \nu_j \left[  \tilde{z}_{ij}^{-(d-2)}   +
 \zeta ( d-2,1-\tilde{z}_{ij})+\zeta ( d-2,1+\tilde{z}_{ij}) \right]
 \Big) \ .  \label{obelambda}
\end{eqnarray}
Here $
\zeta(s,1+a)= \sum_{m=1}^{\infty}(m+a)^{-s}$
is the Generalized Riemann Zeta function and $\tilde{z}_{ij} \equiv z_{ij}/L$
labels the distance in the $z$ direction between the $j^{\rm th}$ and $i^{\rm
th}$ black hole (see Eq.~(2.24) of \cite{obeDias:2007hg} for precise definitions).

Using now \eqref{obelim F} with \eqref{obePhi} one obtains
the behavior of the Newtonian potential $\Phi$ near the $i^{\rm th}$
black hole. This shows that the first term in
\eqref{obelim F} corresponds to the flat space gravitational potential due to the
$i^{\rm th}$ mass $M_i = \nu_i M$ and the second term is a constant potential
due to its images and the presence of the other masses and their images.
The quantity $\Lambda^{(i)}$ plays a crucial role in the explicit construction
of the first-order corrected metric of multi-black holes configurations
on the cylinder and also enters the first-order corrected thermodynamics
(see Sec.~\ref{obesec:ther}).

\runinhead{Equilibrium conditions.}
The third term in \eqref{obelim F} is proportional to $\rho \cos \theta = z
- z_i^*$ and therefore this term  gives a
non-zero constant term in $\partial_z \Phi$ if $\Lambda_1^{(i)}$ is non-zero.
This therefore corresponds to the external force on the $i^{\rm th}$
black hole, due to the other $k-1$ black holes. Indeed, $\Lambda_1^{(i)}$ can be written
as a sum of the potential gradients corresponding to the gravitational force due to each of
the $k-1$ other black holes on the $i^{\rm th}$ black hole as
\begin{eqnarray}
\Lambda_1^{(i)} = \sum_{\substack{j=1,j\neq i } }^k  \nu_j V_{i j}
\,,
  \label{obeLambdaPotent}
\end{eqnarray}
where $V_{ij}$ corresponds to the gravitational field on the $i^{\rm
th}$ black hole from the $j^{\rm th}$ black hole, given by
\begin{eqnarray}
V_{ij} =  \frac{(d-2)}{L^{d-1}}  \left\{ \tilde{z}_{ij}^{-(d-1)} -
 \zeta ( d-1,1-\tilde{z}_{ij} ) +
\zeta (d-1,1+\tilde{z}_{ij} )  \right\} \,,
  \label{obeVij}
\end{eqnarray}
for $j \neq i$. Defining $F_{ij} \equiv \nu_i
\nu_j V_{ij}$ as the Newtonian force on the $i^{\rm th}$ mass due to
the $j^{\rm th}$ mass (and its images as seen in the covering space
of the circle), the condition $\Lambda_1^{(i)}=0$ can be written
as the condition of zero external force on each of the $k$ masses
\begin{equation}
\label{obenoforce} \sum_{j=1,j \neq i}^k F_{ij} = 0  \ ,
\end{equation}
for $i=1,...,k$. As a check, note that it is not difficult to verify that  Newton's law $F_{ij} = - F_{ji}$
is verified using an appropriate identity
for the Generalized Zeta function (see Eq.~(3.6) of \cite{obeDias:2007hg}).

We thus conclude that for static solutions one needs to require the equilibrium condition
$\Lambda_1^{(i)}=0$ for all $i$, since otherwise the $i^{\rm th}$
black hole would accelerate along the $z$ axis.
This gives conditions on the
relation between the positions $z_i^*$ and the mass ratios $\nu_i$, which
are examined in detail in Ref.~\cite{obeDias:2007hg}.
It is shown how to build such equilibrium configurations and
a general copying mechanism is described that builds new
equilibrium configurations by copying any given equilibrium
configuration a number of times around the cylinder.

Note that this equilibrium is an unstable equilibrium, \ie generic small
disturbances in the position of one of the black holes will
disturb the balance of the configuration and result in
the merger of all of the black holes into a single black hole.
As also argued in Ref.~\cite{obeDias:2007hg}, it is expected that
these equilibrium conditions are a consequence of regularity of the solution
since with a non-zero Newtonian force present on the black hole the only way
to keep it static is to introduce a counter-balancing force supported by
a singularity. It turns out that the irregularity of the solution cannot be seen
at the leading order since the binding energy, which accounts for
the self-interaction of the solution, is neglected. It is therefore expected that
singularities will appear at the second order in the total mass for solutions that do no
obey the equilibrium condition mentioned above.

\subsection{Newtonian derivation of the thermodynamics \label{obesec:ther} }

It turns out that there is a quick route to determine the first-order corrected
thermodynamics of the multi-black hole configurations, as explained
in Ref.~\cite{obeDias:2007hg} following the method first found in Ref.~\cite{obeGorbonos:2005px}.
Here one assumes the equilibrium condition \eqref{obenoforce} to be satisfied
and all one needs is the quantity $\Lambda^{(i)}$ defined in \eqref{obelambda},
\ie one does not need to compute the first-order corrected metric.

To start, we define for each black hole an ``areal'' radius
 $\hat{\rho}_{0(i)}$, $i=1, \ldots, k$, such that the individual
 mass, entropy and temperature of each black hole are given by
 \begin{equation}\label{obeSi}
M_{0(i)} =
\frac{(d-1)\Omega_{d-1}}{16\pi G}\hat{\rho}_{0(i)}^{d-2} \spa
 S_{0(i)} = \frac{\Omega_{d-1}  \hat{\rho}_{0(i)}^{d-1}}{4 G} \spa T
_{0(i)} = \frac{d-2}{4\pi\hat{\rho}_{0(i)}} \  .
\end{equation}
These are the intrinsic thermodynamic quantities associated to  each black hole
when they would be isolated in flat empty $(d+1)$-dimensional space.

If we now imagine placing the black holes on a circle at locations $z_i^*$ each
of them will experience a gravitational potential $\Phi_i$. In particular, this
is the Newtonian potential created by all images of
the  $i^{\rm th}$ black hole as well as all other $k-1$ masses (and
their images)  as seen from the location of the $i^{\rm th}$ black
hole. It is not difficult to show that $\Phi_i$ is given by
\begin{equation}
\label{obePhii}
\Phi_i = - \frac{\Lambda^{(i)}}{2 \nu_i} \hat{\rho}_{0(i)}^{d-2} \ ,
\end{equation}
in terms of $\Lambda^{(i)}$  defined in Eq.~\eqref{obelambda}.
Taking into account this potential, we can now determine the
thermodynamic quantities of the interacting system to leading order. By definition,
the entropy $S_i=S_{0(i)}$ is unchanged. The temperature of each black hole, however,
receives a redshift contribution coming from the gravitational potential
$\Phi_i$, so that
\begin{equation}
\label{obetempi}
T_i = T_{0(i)} ( 1 + \Phi_i ) \ .
\end{equation}
The total mass of the configuration is equal to the sum of the individual
masses when the black holes would be isolated
plus the negative gravitational (Newtonian) potential energy
that appears as a consequence of the black holes and their images.
We thus have that the total mass is given by
\begin{equation}
\label{obeMareal} M =M_{0} +U_{\rm Newton} \ ,
\end{equation}
where
\begin{equation}
M_{0} \equiv \sum_{i=1}^k  M_{0(i)} \spa U_{\rm Newton} \equiv
\frac{1}{2}\sum_{i=1}^k  M_{0(i)} \Phi_i \ .
\end{equation}

From these Newtonian results one can then derive the formula
for the relative tension simply by using the (generalized)
first law of thermodynamics (see Eq.~\eqref{obefirstlaw})
\begin{equation}
\delta M = \sum_{i=1}^k T_i \delta S_i + \frac{n M }{L} \delta L \ ,
\end{equation}
from which one finds
\begin{equation}
\label{obender}
n = \frac{L}{M} \left( \frac{ \partial M}{\partial L} \right)_{S_i} \ .
\end{equation}
The condition of keeping $S_i$ fixed means that we should keep fixed the mass
$M_{0(i)}$ of each black hole, and hence also the total intrinsic mass $M_0$.
It this follows from \eqref{obender} and \eqref{obeMareal} that
\begin{equation}
\label{obender2}
n = \frac{L}{M_0} \left( \frac{ \partial U_{\rm Newton}}{\partial L} \right)_{M_{0(i)}}
= -\frac{1}{4M_0} \sum_{i=1}^k M_{0(i)} \frac{\hat{\rho}_{0(i)}^{d-2}}{\nu_i}   L
\frac{ \partial \Lambda^{(i)}}{\partial L}  = \frac{d-2}{4}
\sum_{i=1}^k  \Lambda^{(i)}\hat{\rho}_{0(i)}^{d-2}  \ ,
\end{equation}
where we used that $\Lambda^{(i)} \propto L^{-(d-2)}$ for fixed locations $z_i^*$
(see Eq.~\eqref{obelambda}) and $M_{0(i)} = \nu_i M_0$.
As shown in \cite{obeDias:2007hg}, the thermodynamics above agrees with the explicitly
computed thermodynamic quantities from the first-order corrected metric.

We emphasize that these results are correct only to first order in the mass
and note that in terms of the reduced mass \eqref{obethemu} the expression \eqref{obender2}
gives that $n$ as a function of $\mu$ is given for the multi-black hole
configurations by
\begin{eqnarray}
\label{obenofmu2} n (\mu)  = \frac{(d-2)(2\pi)^{d-2} }{4(d-1) \Omega_{d-1}}
\sum_{i=1}^k \nu_i \Lambda^{(i)} \mu + \CO (\mu^2) \ ,
\end{eqnarray}
which generalizes the single-black hole result given in \eqref{obenofmu}.
In terms of the phase diagram of Fig.~\ref{obefig1}, it follows from this result
that (at least for small masses) the $k$ black hole configurations correspond
to points lying above the single-black hole phase and below the $k$ copied phase.

From the first-order corrected temperatures \eqref{obetempi} one can show that the
multi-black hole configuration are in general not in thermal
equilibrium. The only configurations that are in thermal equilibrium
to this order are the copies of the single-black hole solution
studied previously \cite{obeHorowitz:2002dc,obeHarmark:2003eg,obeHarmark:2003yz}.
As a further comment we note that Hawking radiation will seed the mechanical
instabilities of the multi-black hole configurations.
The reason for this is that in a generic configuration the black
holes have different rates of energy loss and hence the mass ratios required
for mechanical equilibrium are not maintained.
This happens even in special configurations, \eg when the temperatures are equal,
because the thermal radiation is only statistically uniform. Hence asymmetries
in the real time emission process will introduce disturbances
 driving these special configurations away from their equilibrium positions.

\subsection{Consequences for the phase diagram \label{obesec:copd} }

The existence of the  multi-black hole solutions has striking consequences for
the phase structure of black hole solutions on $\CM^d \times S^1$.
It means that one can for example start from a solution with two
equal size black holes, placed oppositely to each other on the
cylinder, and then continuously deform the solution to be
arbitrarily close to a solution with only one black hole (the other
black hole being arbitrarily small in comparison). Thus, we get a
continuous span of classical static solutions for a given total
mass. In particular, a multi-black hole configuration with $k$ black holes has $k$
independent parameters. This implies a continuous non-uniqueness in
the $(\mu,n)$ phase diagram (or for a given mass), much like the one
observed for bubble-black hole sequences \cite{obeElvang:2004iz}
and for other classes of black hole solutions
\cite{obeEmparan:2004wy,obeElvang:2007rd,obeIguchi:2007is,obeElvang:2007hg} (see also
Sec.~\ref{obesec:phas}). In particular, this has the consequence that if we would live on $\CM^4
\times S^1$ then from a four-dimensional point of view one would
have an infinite non-uniqueness for static black holes of size
similar to the size of the extra dimension, thus severely breaking
the uniqueness of the Schwarzschild black hole.

\runinhead{New non-uniform strings ?}
Another consequence of the new multi-black hole configurations
is for the connection to uniform and non-uniforms strings
on the cylinder. As discussed in Sec.~\ref{obesec:pdia}, there is evidence that the
black hole on the cylinder phase merges with the non-uniform black
string phase in a topology changing transition point. It follows
from this that the copies of black hole on the cylinder solution
merge with the copies of non-uniform black strings. However, due to
the multi-black hole configurations we now
have a continuous span of solutions connected
to the copies of the black hole on the cylinder. Therefore, it is
natural to ask whether the new solutions also merge with non-uniform
black string solutions in a topology changing transition point. If
so, it probes the question whether there exist, in addition to
having new black hole on the cylinder solutions, also new
non-uniform black string solutions. Thus, these new solution
present a challenge for the current understanding of the
phase diagram for black holes and strings on the cylinder. For a detailed
discussion on this see Ref.~\cite{obeDias:2007hg}.

Another connection with strings and black holes on the cylinder
is that a Gregory-Laflamme unstable uniform black string is believed
to decay to a black hole on the cylinder (when the number of
dimensions is less than the critical one \cite{obeSorkin:2004qq}).
However, the new multi-black hole solutions mean that one can imagine
them as intermediate steps in the decay.

\runinhead{Lumpy black holes.}
Ref.~\cite{obeDias:2007hg} also examines in detail configurations with two and three
black  holes. For two black holes this confirms the expectation that
one maximizes the entropy by transferring all the mass to one of the
black holes, and also that if the two black holes are not in mechanical
equilibrium then the entropy is increasing as the black holes become
closer to each other. These two facts are both in accordance with
the general argument that the multi-black hole configurations are in
an unstable equilibrium and generic perturbations of one of the positions
will result in that all the black holes merge together in a single
black hole on the cylinder.

A detailed examination of the  three black hole solution
suggests the possibility of further new types of black hole solutions
in Kaluza-Klein spacetimes. In particular, this analysis
suggests the possibility that new
static configurations may exist that consist of a lumpy black hole,
where the non-uniformities are
supported by the gravitational stresses imposed by an external field.
These new solutions were argued by considering a symmetric configuration
of three black holes, with one of mass $M_1$ and two others of equal mass $M_2=M_3$
at equal distance to the first one. Increasing the total mass of the system
shows that it is possible that the two black holes (2 and 3) merge before
merging with black hole 1. In this way one could end up with a static solution
consisting of lumpy black hole (\ie a `peanut-like' shaped black object)
together with an ellipsoidal  black hole.

\runinhead{Analogue fluid model.}
Finally we note
that one may consider the multi-black hole configurations in relation to an
analogue fluid model for the Gregory-Laflamme (GL) instability,
recently proposed in Ref.~\cite{obeCardoso:2006ks}.
There it was pointed out that the GL instability of a black string has a natural
analogue description in terms of the Rayleigh-Plateau (RP) instability of a fluid cylinder.
It turns out that many known properties of the gravitational  instability
have an analogous manifestation in the fluid model. These include the behavior of
the threshold mode with $d$, dispersion relations, the existence of critical dimensions
and the initial stages of the time evolution
(see Refs.~\cite{obeCardoso:2006ks,obeCardoso:2006sj,obeCardoso:2007ka} for details).
In the context of this analogue fluid model, Ref.~\cite{obeDias:2007hg}
discusses a possible, but more speculative, relation of the
multi-black hole configurations to configurations observed in the
time evolution of fluid cylinders.

\section{Thin black rings in higher dimensions \label{obesec:robh} }

In this and the next section we turn our attention to rotating black holes.
We start by reviewing the recent construction \cite{obeEmparan:2007wm}
of an approximate solution for an asymptotically flat neutral thin rotating
black ring  in any dimension $D \geq 5$ with horizon topology $S^{D-3}\times S^1$.
As in Sec.~\ref{obesec:mubh}, this construction uses the method of matched asymptotic expansion, and we only present the main
points. We discuss in particular the equilibrium condition necessary for
balancing the ring, and how this enables to obtain the leading order thermodynamics
of thin rotating black rings. We also compare the thermodynamics of the thin
black ring to that of the MP black hole. In this and the following section we denote the
number of spacetime dimensions by $D=4+n$.

\subsection{Thin black rings from boosted black strings \label{obesec:boos}}

Black rings in $(n+4)$-dimensional asymptotically flat spacetime are
solutions of Einstein gravity with an event horizon of topology $S^1\times S^{n+1}$.
As we briefly reviewed in Secs.~\ref{obesec:intr} and \ref{obesec:uniq} explicit solutions with
this topology in five dimensions $(n=1)$
were first presented in Ref.~\cite{obeEmparan:2001wn}
(see also \cite{obeEmparan:2006mm} for a review).

In five dimensions, there is beyond the MP black hole and the black ring one more
phase of rotating black holes if one restricts to phases with a single
angular momentum that are in  thermal equilibrium. This is the black Saturn phase
consisting of a central MP black and one black ring around it, having equal
temperature and angular velocity. If one abandons the
condition of thermal equilibrium there are many more black Saturn phases with multiple
rings as well as multi-black ring solutions.
We refer to \cite{obeElvang:2007hg} and the recent review \cite{obeEmparan:2008eg}
for details on the more general phase structure for the five-dimensional case.

The construction of analogous solutions in more than five dimensions
is considerably more involved, since for $D \geq 6$ these solutions
are not contained in the  generalized Weyl ansatz
\cite{obeEmparan:2001wk,obeHarmark:2004rm,obeHarmark:2005vn} because
they do not have $D-2$ commuting Killing symmetries.
Furthermore the inverse scattering techniques of
\cite{obeBelinsky:1971nt,obeBelinsky:1979,obeBelinski:2001ph,obePomeransky:2005sj}
do not extend to the asymptotically flat case in any $D\geq 6$.

Therefore, one way to make progress towards solving this problem can
be achieved by first constructing thin black ring solutions in arbitrary dimensions
as a perturbative expansion around circular boosted black strings.
The idea that rotating thin black rings should be well approximated by boosted
black strings is intuitively clear and already appears in earlier works
\cite{obeElvang:2003mj,obeHovdebo:2006jy,obeElvang:2006dd}. This was
used as a starting point in the explicit construction \cite{obeEmparan:2007wm}.

\runinhead{Boosted black string.}
The zeroth order solution is that of a \textit{straight} boosted black string.
 The metric of this can easily be obtained from \eqref{obeublstr} by applying
a boost in the $(t,z)$ plane. The result is
\begin{eqnarray}
\label{obeappab}
ds^2 &=& -\left( 1 - \cosh^2 \alpha \frac{r_0^{n}}{r^{n}} \right) dt^2 -
 2 \frac{r_0^{n}}{r^{n}}
 \cosh \alpha \sinh \alpha\, dt dz + \left( 1 + \sinh^2 \alpha
\frac{r_0^{n}}{r^{n}} \right) dz^2 \nn \\ &&
 + \left( 1- \frac{r_0^{n}}{r^{n}} \right)^{-1} dr^2 + r^2 d\Omega_{n+1}^2 \,,
\end{eqnarray}
where $r_0$ is the horizon radius and $\alpha$ is the boost parameter.
In general, we will take the $z$ direction to be along an $S^1$ with
circumference $2\pi R$, which means we can write $z$ in terms of an
angular coordinate $\psi$ defined by $ \psi=z/R$ ($ 0\leq \psi <2\pi$).
At distances $r\ll R$, the solution \eqref{obeappab} is the approximate
metric of a thin black ring to zeroth order in $1/R$.

By definition, a thin black ring has an $S^1$
radius $R$ that is much larger than its $S^{n+1}$ radius $r_0$. In this limit,
the mass of the black ring is small and the gravitational attraction
between diametrically opposite points of the ring is very weak. So,
in regions away from the black ring, the linearized approximation to
gravity will be valid, and the metric will be well-approximated if we
substitute the ring by an appropriate delta-like distributional source
of energy-momentum. The source has to be chosen so that the metric it produces
is the same as that expected from the full exact solution in the region
far away from the ring. Since the thin black ring is expected to
approach locally the solution for a boosted black string, it is sensible
to choose distributional sources that reproduce the metric \eqref{obeappab}
in the weak-field regime,
\begin{subequations}
\label{obesource}
\begin{eqnarray}
T_{tt}&=&\frac{r_0^{n}}{16\pi G}\,\left(n\cosh^2\alpha+1\right)\,\delta^{(n+2)}(r)\,,\\
\label{obedistsourcea}
T_{tz}&=&\frac{r_0^{n}}{16\pi G}\,n\cosh\alpha\sinh\alpha\,\delta^{(n+2)}(r)\,,\\
\label{obedistsourceb}
T_{zz}&=&\frac{r_0^{n}}{16\pi G}\,\left(n\sinh^2\alpha-1\right)\,\delta^{(n+2)}(r)\,.
\label{obedistsourcec}
\end{eqnarray}
\end{subequations}
The location $r=0$ corresponds to
a circle of radius $R$ in the $(n+3)$-dimensional Euclidean flat space,
parameterized by the angular coordinate $\psi$. In this construction the
mass  and angular momentum
of the black ring are obtained by integrating the energy and momentum densities,
\begin{equation}
\label{obeMJa}
M=2\pi R \int_{S^{n+1}} T_{tt} \spa
J =2\pi R^2 \int_{S^{n+1}} T_{tz} \ ,
\end{equation}
where $S^{n+1}$ links the ring once.

\runinhead{Dynamical equilibrium condition.}
We now first show that the boost parameter $\alpha$ gets
fixed by a  dynamical equilibrium condition ensuring that the string tension is
balanced against the centrifugal repulsion. To this end note that we
are approximating the black ring by a distributional source of energy-momentum.
The general form of the equation of motion for probe brane-like objects
in the absence of external forces takes the form \cite{obeCarter:2000wv}
\begin{equation}
\label{obeKT}
{K_{\mu\nu}}^{\rho}T^{\mu\nu}=0\,,
\end{equation}
where the indices $\mu,\nu$ are tangent to the brane and $\rho$ is
transverse to it. The second fundamental tensor
${K_{\mu\nu}}^{\rho}$ extends the
notion of extrinsic curvature to submanifolds of codimension possibly larger than
one. The extrinsic curvature of the circle is $1/R$, so a circular
linear distribution of energy-momentum of
radius $R$ will be in equilibrium only if
\begin{equation}
\label{obenotzz}
\frac{T_{zz}}{R}=0\,,
\end{equation}
\ie for finite radius the pressure tangential to the circle must
vanish. Hence, for the thin black ring with source \eqref{obesource},
 the condition that the ring be in equilibrium translates into a very specific value
for the boost parameter
\begin{equation} \label{obeeqboost}
\sinh^2\alpha=\frac{1}{n}\,,
\end{equation}
which we will also refer to as the critical boost.
For $D=5$ ($n=1$) this was already observed in Ref.~\cite{obeElvang:2003mj}
where the thin black string limit of
five-dimensional black rings was first made explicit, but
the connection with \eqref{obeKT} was first noticed in \cite{obeEmparan:2007wm}.

\runinhead{Thermodynamics.}
Using \eqref{obeeqboost} it is not difficult to obtain the physical quantities
of the critically boosted black string, and hence the leading-order thermodynamics
of thin black rings (see also Refs.~\cite{obeHovdebo:2006jy,obeKastor:2007wr}
for further details on boosted black strings and their thermodynamics).
We find for the mass $M$, entropy $S$, temperature $T$,
angular momentum $J$ and angular velocity $\Omega$ the expressions \cite{obeEmparan:2007wm}
\begin{subequations}
\label{obetbrthermo}
\begin{equation}
\label{obetbrthermo1}
M=\frac{\Omega_{n+1}}{8 G}\,R\, r_0^{n}(n+2) \spa
S=\frac{\pi\,\Omega_{n+1}}{2G}  R\,r_0^{n+1}
\sqrt{\frac{n+1}{n}} \spa T = \frac{n}{4\pi} \sqrt{
\frac{n}{n+1}} \frac{1}{r_0}\,,
\end{equation}
\begin{equation}
\label{obetbrthermo2}
J=\frac{\Omega_{n+1}}{8 G}\,R^2\, r_0^{n}\sqrt{n+1}\,,\qquad
\Omega = \frac{1}{\sqrt{n+1}} \frac{1}{R}\,.
\end{equation}
\end{subequations}
We also note that an equivalent but more physical form of the
equilibrium equation \eqref{obeeqboost} in terms of these quantities is
\begin{equation}
\label{obeRJM}
R=\frac{n+2}{\sqrt{n+1}}\frac{J}{M}\,.
\end{equation}
We thus see that the radius grows linearly with $J$ for fixed mass.

It is remarkable that with the above reasoning one can already obtain
the correct limiting thermodynamics of thin black rings to leading order,
without having to solve for any metric. One finds from \eqref{obetbrthermo}
that the entropy of thin black rings behaves as
\begin{equation}
\label{obesmjring}
S^{\rm ring}(M,J) \propto J^{-\frac{1}{D-4}}\;M^{\frac{D-2}{D-4}}
\,,
\end{equation}
whereas that of ultra-spinning MP black holes in $D \geq 6$ is given by
\cite{obeEmparan:2003sy}
\begin{equation}
\label{obesmjhole}
S^{\rm hole} (M,J) \propto J^{-\frac{2}{D-5}}\;M^{\frac{D-2}{D-5}}
\,.
\end{equation}
This already shows the non-trivial fact that in the ultra-spinning regime of large $J$ for
fixed mass $M$ the rotating black ring has higher entropy than the MP
black hole (see also Sec.~\ref{obesec:mpbh}).
Moreover, as will be explained in Sec.~\ref{obesec:maex},
it turns out that for $D \geq 6$ the results \eqref{obetbrthermo} are actually
valid up to and including the next order in $r_0/R$, so receives only
$O(r_0^2/R^2)$ corrections. This conclusion could already be drawn once one
has convinced oneself that the first-order $1/R$  correction terms in the metric
only involve dipole contributions which can easily be argued to give zero
contribution to all thermodynamic quantities \cite{obeEmparan:2007wm}.

It is important to stress that the above reasoning relies crucially on the
assumption that when the boosted black string is curved, the horizon remains regular.
To verify this point, and also to obtain a metric for the thin black ring,
Ref.~\cite{obeEmparan:2007wm} solves the Einstein equations explicitly by
constructing an approximate solution for $r_0\ll R$ using a
matched asymptotic expansion. In this analysis one finds that
the condition \eqref{obenotzz} appears as a consequence of demanding
absence of singularities on the plane of the ring outside
the horizon. Whenever $n\sinh^2\alpha\neq 1$ with finite $R$,
the geometry backreacts creating singularities on the plane of the ring.
These singularities admit a natural interpretation. Since \eqref{obeKT} is a
consequence of the conservation of the energy-momentum tensor, when
\eqref{obenotzz} is not satisfied there must be additional sources of
energy-momentum. These additional sources are responsible
for the singularities in the geometry.
Alternatively, the derivation of \eqref{obenotzz} in Ref.~\cite{obeEmparan:2007wm}
from the Einstein equations is an example of how General Relativity encodes the equations of motion of
black holes as regularity conditions on the geometry.

\subsection{Matched asymptotic expansion \label{obesec:maex} }

We now review the highlights of the perturbative construction of
thin black rings using matched asymptotic expansion (see also
Sec.~\ref{obesec:cons}). In the problem at hand, the two widely separated
scales are the `thickness' of the ring $r_0$ and the radius of the
ring $R$, and the thin limit means that $r_0 \ll R$.
There are therefore two zones,   an asymptotic zone at large distances from the
black ring, $r\gg r_0$, where the field can be expanded in powers of $r_0$.
The other zone is the near-horizon zone which lies at scales much smaller than the
ring radius, $r\ll R$. In this zone the field is expanded in powers of $1/R$. At
each step, the solution in one of the zones is used to provide boundary
conditions for the field in the other zone, by matching the fields in
the `overlap' zone $r_0\ll r \ll R$ where both expansions are valid.

As already discussed in Sec.~\ref{obesec:boos}, the starting point is to
consider the solution in the near-horizon zone to zeroth order in $1/R$,
 \ie we take a boosted black string of infinite length, $R\to\infty$.
 The next steps in the construction are then as follows:

\begin{itemize}
\item Step 1: One solves the Einstein equations in the linearized
approximation around flat space for a source corresponding to
 a circular distribution of a given mass and momentum density as given
in \eqref{obesource}.  This metric is valid in the region  $r \gg r_0$.
\item Step 2: We consider the Newtonian solution close to the sources, \ie
in the overlap region $r_0\ll r\ll R$.
\item Step 3: We consider the near-horizon region of the ring and
 find the linear corrections to the metric of a boosted black
string for a perturbation that is small in $1/R$; in other words, we
analyze the geometry of a boosted black
string that is now slightly curved into a circular shape. This solution is then
matched to the metric in the overlap region found in Step 2. The resulting
solution is valid in the region $r_0 \leq r \ll L$.
\end{itemize}

To solve Step 1 for a non-zero $T_{\psi\psi}=R^2\, T_{zz}$ is not easy. It
is therefore convenient to already assume that the equilibrium condition $T_{\psi\psi}=0$
in \eqref{obenotzz} is satisfied. This then gives the solution of a black ring in
linearized gravity \cite{obeEmparan:2007wm}.
Finding a more general solution with a source for the tension is much easier
if one restricts to the overlap zone (Step 2). In this regime we are studying the effects
of locally curving a thin black string into an arc of constant curvature radius
$R$. To this end it is convenient to introduce ring-adapted coordinates.
These are derived in Ref.~\cite{obeEmparan:2007wm} and to first-order in $1/R$
the flat space metric in these coordinates takes the form
\begin{equation}\label{obeadapted}
ds^2(\bbe{n+3}) =\left( 1 + \frac{2r\cos \theta}{R}
\right) dz^2 +\left( 1 - \frac{2}{n}\frac{r \cos \theta}{R} \right)
\left( dr^2 + r^2 d\theta^2 + r^2 \sin^2 \theta d\Omega_{n}^2
\right)\, .
\end{equation}
In terms of these coordinates the general form of the metric in the overlap
region is then
\begin{equation}
\label{obecorover}
g_{\mu \nu} \simeq \eta_{\mu \nu} + \frac{r_0^n}{r^n} \left( h_{\mu \nu}^{(0)} (r) +
\frac{r \cos \theta}{ R} h_{\mu \nu}^{(1)} (r) \right) \ .
\end{equation}
Solving Einstein equations to order $1/R$ then explicitly shows that regularity
of the solution enforces vanishing of the tension $T_{zz}$ (see Eq.~\eqref{obenotzz}).

The technically most difficult part of the problem is to find the near-horizon solution
in step 3. Physically, this corresponds  to curving the black string into a circle of large but
finite radius $R$. In effect, this means that we are placing the black string in
an external potential whose form at large distances is that of \eqref{obecorover}
and which changes the metric $g_{\mu \nu}^{\rm bbs}$ in Eq.~\eqref{obeappab} of the
(critically) boosted black string by a small amount, \ie.
\begin{equation}
\label{obecorrmet}
g_{\mu \nu} \simeq g_{\mu \nu}^{\rm bbs} (r;r_0) + \frac{\cos \theta}{R}
h_{\mu \nu} (r;r_0) \ .
\end{equation}
In Ref.~\cite{obeEmparan:2007wm} the Einstein equations to order $1/R$ are explicitly
solved, showing that
the perturbations $h_{\mu \nu} (r;r_0)$ can be expressed in terms of hypergeometric functions.

\runinhead{Corrected thermodynamics.}
One can find the corrections to the thermodynamics as follows.
First, one uses the near-horizon corrected metric \eqref{obecorrmet} to  find
the corrections to the entropy $S$, temperature $T$, and angular
velocity $\Omega$. Then one can use the 1st law,
\begin{equation}
\label{obefirstlawa}
dM= T \delta S +\Omega \delta J \ ,
\end{equation}
and the Smarr formula
\begin{equation}
\label{obesmarr}
(n+1)M=(n+2) \left( T S+\Omega J \right) \ ,
\end{equation}
to deduce the corrections to the mass and angular momentum.%
\footnote{This method was also used in Ref.~\cite{obeHarmark:2003yz,obeDias:2007hg}
for small black holes and multi-black holes on the cylinder.}
Using now that the perturbations in \eqref{obecorrmet} are only of dipole type,
with no monopole terms, it follows that the area, surface gravity and angular
velocity receive no modifications in $1/R$. The reason is that a
dipole can not change the total area of the horizon, only its shape.
This is true both of the shape of the $S^{n+1}$ as well as the length of
the $S^1$, which can vary with $\theta$ but on average (\ie when
integrated over the horizon) remains constant. So $S$ is not
corrected. The surface gravity and angular velocity can not be corrected
either. They must remain uniform on a regular horizon, so, since the
dipole terms vanish at $\theta=\pi/2$, no corrections to $T$
and $\Omega$ are possible. It then follows from
\eqref{obefirstlawa}, \eqref{obesmarr} that $M$ and $J$ are not corrected either.%
 \footnote{In
five-dimensions ($n=1$) there {\em are} corrections to this order. Their
origin is discussed in App.~A of \cite{obeEmparan:2007wm}.} So the function
$S(M,J)$ obtained in \eqref{obesmjring} is indeed valid including the
first order in $1/R$. It is interesting to observe that this conclusion
could be drawn already when the asymptotic form of the metric \eqref{obecorover}
in the overlap zone, is seen to include only dipole terms at order $1/R$.

\subsection{Black rings versus MP black holes \label{obesec:mpbh} }

We now proceed by  analyzing the thin black ring thermodynamics
and compare it to that of ultra-spinning MP black holes. Recall that
the thermodynamics of the thin black ring in the ultra-spinning regime is given
by \eqref{obetbrthermo}, which is valid up to $O(r_0^2/R^2)$ corrections.

\runinhead{Myers-Perry black hole.}
For the  MP black hole, exact results can be obtained for all values of
the rotation.
 The two independent parameters specifying the (single-angular momentum)
 solution are the mass parameter $\mu$ and the rotation parameter $a$, from which the horizon
radius $r_0$ is found as the largest (real) root of the
equation
\begin{equation}\label{obemueq}
\mu = (r_0^2 + a^2) r_0^{n-1}\,.
\end{equation}
In terms of these parameters the thermodynamics take
the form \cite{obeMyers:1986un}
\begin{subequations}\label{obeTMPJOm}
\begin{equation}
\label{obeTMP}
 M = \frac{ (n+2)  \Omega_{n+2} \, \mu}{16 \pi G}
\,,\qquad  S = \frac{\Omega_{n+2} \, r_0 \, \mu}{4 G} \,,\qquad
T = \frac{1}{4 \pi} \left( \frac{2 r_0^n}{\mu} + \frac{n-1}{r_0}
\right)\,,
\end{equation}
\begin{equation}\label{obeJOm}
J = \frac{ \Omega_{n+2}\, a \, \mu }{ 8 \pi G}
\,,\qquad \Omega = \frac{a \, r_0^{n-1}}{\mu}\,.
\end{equation}
\end{subequations}
Note the similarity between $a=\frac{n+2}{2} \frac{J}{M}$
and the black ring relation \eqref{obeRJM}.

An important simplification occurs in the ultra-spinning regime
of $J\to\infty$ with fixed $M$, which corresponds to
$a \rightarrow \infty$. Then \eqref{obemueq} becomes $ \mu \rightarrow a^2 r_0^{n-1}$
leading to simple expressions for Eq.~\eqref{obeTMPJOm} in terms of $r_0$ and
$a$, which in this regime play roles analogous to those of $r_0$ and $R$ for the
black ring. Specifically, $a$ is a measure of the size of the horizon along the
rotation plane and $r_0$ a measure of the size transverse
to this plane \cite{obeEmparan:2003sy}. In fact, in this limit
\begin{equation}
\label{obeTMP2}
 M \to \frac{ (n+2)  \Omega_{n+2}}{16 \pi G}\; a^2 r_0^{n-1}
\,,\qquad S \to \frac{\Omega_{n+2}}{4 G}\;a^2 r_0^{n} \,,\qquad
T \to \frac{n-1}{4 \pi r_0} \ ,
\end{equation}
take the same form as the expressions characterizing a black membrane extended
along an area $\sim a^2$ with horizon radius $r_0$.
This identification lies at the core of the ideas in \cite{obeEmparan:2003sy},
which were further developed in Ref.~\cite{obeEmparan:2007wm}
and will be summarized in Sec.~\ref{obesec:phas}.
We note that the quantities $J$ and $\Omega$ disappear since
the black membrane limit is approached in the region near the axis of rotation of
the horizon and so  the membrane is static in the limit. Note furthermore that
\eqref{obeTMP2} is valid up to $O(r_0^2/a^2)$ corrections.

Finally, we remark that the transition to the membrane-like regime is signaled by a
qualitative change in the thermodynamics of the MP black holes.
At $a/r_0 = \sqrt(\frac{n+1}{n-1})$ the temperature reaches a minimum and
$\left(\partial^2 S/\partial J^2\right)_M$
changes sign. For $a/r_0$ smaller than this value, the thermodynamic quantities
of the MP black holes such as $T$ and $S$ behave similarly to those of the Kerr
solution and one should not expect any membrane-like behavior.
However, past this point they rapidly approach the membrane results.
We do not expect that the onset of thermodynamic instability at this point
is directly associated to any dynamical instability. Rather, one expects
a GL-like instability to happen at a larger value of $a/r_0$
\cite{obeEmparan:2003sy,obeEmparan:2007wm}.

\runinhead{Dimensionless quantities.}
Contrary to the case of KK black holes where we could use the circle length
to define dimensionless quantities (cf. \eqref{obethemu} or \eqref{obeeladef}) in this
case we need to use one of the physical parameters of the solutions to define dimensionless
quantities. We choose the mass $M$ and thus introduce dimensionless quantities for the
spin $j$, the area $a_H$, the angular
velocity $\omega_H$ and the temperature $\mathfrak{t}_H$ via
\begin{subequations}\label{obejaot}
\begin{equation}
\label{obejaHdef}
j^{n+1} \propto \frac{J^{n+1}}{GM^{n+2}} \,,\qquad
a_H^{n+1} \propto  \frac{S^{n+1}}{(GM)^{n+2}}
~,
\end{equation}
\begin{equation}
\label{obeotdef}
\omega_H \propto  \Omega
(GM)^{\frac{1}{n+1}} \,,\qquad
\mathfrak{t}_H  \propto (GM)^{\frac{1}{n+1}}\, T\,,
\end{equation}
\end{subequations}
where convenient normalization factors can be found in Eq.~(7.9) of \cite{obeEmparan:2007wm}.
We take $j$ as our control parameter and now study and compare the
functions $a_H (j)$, $\omega_H(j)$ and $\mathfrak{t}_H (j) $ for black rings and MP black holes
in the ultra-spinning regime. These asymptotic phase curves can now obtained
using \eqref{obejaot} together with  \eqref{obetbrthermo} and \eqref{obeTMP2} respectively.
In the following we denote
the results for the thin black ring with $^{(r)}$ and for the ultra-spinning MP black
holes with $^{(h)}$, and generally omit numerical prefactors.

\begin{figure}[ht]
\begin{picture}(0,0)(0,0)
\put(400,13){$j$}
\put(55,230){$a_H$}
\put(77,3){$j_{\rm mem}$}
\put(123,3){$j_{\rm GL}$}
\end{picture}
\centerline{\includegraphics[width=13cm]{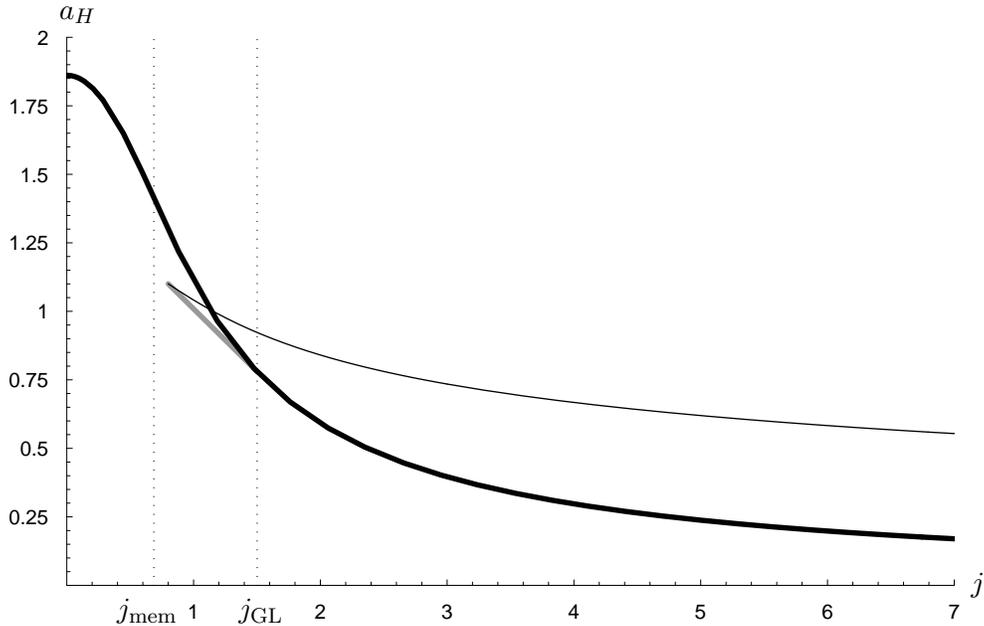}}
\caption{\small Area vs spin for fixed mass, $a_H(j)$, in seven
dimensions. For large $j$, the thin curve is the result for thin black rings
and is extrapolated here down to $j\sim O(1)$. The thick
curve is the exact result for the MP black hole. The gray line corresponds
to the conjectured phase of pinched black holes (see Sec.~\ref{obesec:phas}), which branch off tangentially from the MP
curve at a value $j_{\rm GL}> j_{\rm mem}$.
At any given dimension, the phases should not necessarily
display the swallowtail as shown in this diagram, but could also
connect more smoothly  via a pinched black hole phase that starts tangentially
in $j_{\rm GL}$ and has increasing $j$.  Reprinted from Ref.~\cite{obeEmparan:2007wm}.}
\label{obefig:sevendphases}
\end{figure}

\runinhead{Comparison of the thermodynamics.}
Starting with the reduced area function we see that
\begin{equation}
\label{obeaHrh}
a_H^{(r)}\sim \frac{1}{{j}^{1/n}}\,,\qquad a_H^{(h)}\sim \frac{1}{{j}^{2/(n-1)}}\,,
\end{equation}
and so, for any $D=4+n\geq 6$, the area decreases faster for
MP black holes than for black rings, so we immediately see that black rings dominate
entropically in the ultra-spinning regime \cite{obeEmparan:2007wm}.
For illustration, Fig.~\ref{obefig:sevendphases} shows these curves in $D=7$ ($n=3$).

Including prefactors one finds for the angular velocities that
\begin{equation}
\omega_H^{(r)} \to \frac{1}{2j} \spa \omega_H^{(h)} \to \frac{1}{j} \ .
\end{equation}
The ratio $\omega^{(h)}_H/\omega^{(r)}_H=2$, which holds for all $D\geq
6$, is reminiscent of the factor of 2 in Newtonian mechanics between the
moment of inertia of a wheel (\ie a ring) and a disk (\ie a pancake) of
the same mass and radius, which implies that the disk must rotate twice
as fast as the wheel in order to have the same angular momentum. Irrespective of
whether this is an exact analogy or not, the fact that $\omega^{(r)}_H<\omega^{(h)}_H$
is clearly expected from this sort of picture. For the temperatures we find
\begin{equation} \label{obetempsa}
\mathfrak{t}^{(r)}_H \sim j^{1/n} \spa \mathfrak{t}^{(h)}_H \sim
j^{2/(n-1)}\, , \end{equation}
so the thin black ring is colder than the MP
black hole. In fact, since the temperature is inversely proportional to the thickness
of the object the picture suggested above leads to the
following argument: if we put a given mass in the shape of a wheel
of given radius, then we get a thicker object than if we put it in
the shape of a pancake of the same radius.

\section{Completing the phase diagram \label{obesec:phas} }

In this section we will discuss the phase structure of asympotically flat neutral
rotating black holes in six and higher dimensions by exploiting a connection between,
on one side, black
holes and black branes in KK spacetimes and, on the other side, higher-dimensional
rotating black holes. Building on the basic idea in \cite{obeEmparan:2003sy}, this
phase structure was recently proposed in Ref.~\cite{obeEmparan:2007wm}. Part of this
picture is conjectural, but is based on well-motivated analogies and
appears to be natural from many points.

The curve $a_H(j)$ at values of $j$ outside the domain of validity of
the computations in Sec.~\ref{obesec:robh} corresponds to the regime where
the gravitational self-attraction of the
ring is important. There are no analytical methods presently known to
treat such values $j\sim O(1)$, and the precise form of the curve in this
regime may require numerical solutions. However, as argued in Ref.~\cite{obeEmparan:2007wm}
it is possible to complete the black ring curve and other features of the phase diagram,
at least qualitatively. This is done by combining a number of observations and
reasonable conjectures about the behavior of MP black holes at large rotation and
using as input the presently known phase structure of Kaluza-Klein black holes
(see Sec.~\ref{obesec:kkbh}).

\subsection{GL instability of ultra-spinning MP black hole \label{obesec:usgl}}

In the ultra-spinning regime in $D\geq 6$, MP black holes
approach the geometry of a black membrane $\approx  \R^2 \times
S^{D-4}$ spread out along the plane of rotation \cite{obeEmparan:2003sy}.
In Sec.~\ref{obesec:mpbh} we have already
observed that the extent of the black hole along the plane is approximately
given by the rotation parameter $a$, while the `thickness' of the
membrane, \ie the size of its $S^{D-4}$, is given by the parameter $r_0$.
For $a/r_0$ larger than a critical value of order one
we expect that the dynamics of these black holes
is well-approximated by a black membrane compactified on a square torus
$\T^2$ with side length $L\sim a$ and with $S^{D-4}$ size $\sim r_0$.
The angular velocity of the black hole is always moderate, so it will
not introduce large quantitative
differences, but note that the rotational axial symmetry of the MP black
holes translates into only one translational symmetry along the $\T^2$,
the other one being broken.

\begin{figure}[ht]
\centerline{\includegraphics[width=9.8cm]{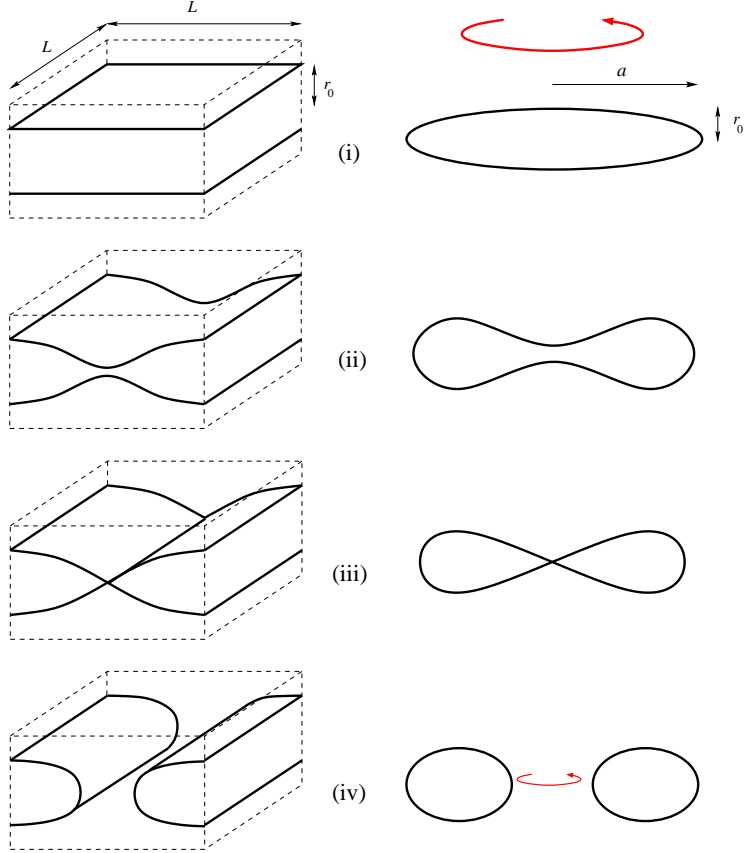}}
\caption{\small Correspondence between phases of black membranes wrapped
on a $\T^2$ of side $L$ (left) and fastly-rotating MP black holes with rotation
parameter $a\sim L\geq r_0$ (right: must be rotated along a vertical axis):
(i) Uniform black membrane and MP black hole.
(ii) Non-uniform black membrane and pinched black hole.
(iii) Pinched-off membrane and black hole. (iv) Localized black string
and black ring. Reprinted from Ref.~\cite{obeEmparan:2007wm}. }
\label{obefig:membranes}
\end{figure}

Using this analog mapping of membranes and fastly rotating MP black
holes, Ref.~\cite{obeEmparan:2003sy} argued that the latter should exhibit
a Gregory-Laflamme-type instability. Furthermore, as reviewed in
Sec.~\ref{obesec:kkbh} it is known that
the threshold mode of the GL instability gives rise to a new branch of
static non-uniform black strings and branes
\cite{obeGregory:1988nb,obeGubser:2001ac,obeWiseman:2002zc}. In correspondence
with this, Ref.~\cite{obeEmparan:2003sy} argued that it is natural to
conjecture the existence of new branches of axisymmetric `lumpy' (or
`pinched') black holes, branching off from the MP solutions along the
stationary axisymmetric zero-mode perturbation of the GL-like
instability,

\runinhead{Map to phases of KK black holes on the torus.}
In Ref.~\cite{obeEmparan:2007wm} this analogy was pushed further by drawing
a correspondence between the phases of KK black holes on the torus (see Sec.~\ref{obesec:torp})
and the phases of higher-dimensional black holes, as illustrated
in Fig.~\ref{obefig:membranes}. Here we have restricted to non-uniformities of the
membrane along only of the two brane directions, since including non-uniformity
in a second direction would not have a counterpart for rotating black holes.
These would break axial symmetry
and hence would be radiated away. Other limitations of the analogy are
discussed in detail in Ref.~\cite{obeEmparan:2007wm}.

Using the correspondence between the phases of the two systems, one can
import, at least qualitatively, the known phase diagram of black
membranes on $\CM^{D-2} \times \T^2$ onto the phase diagram of rotating
black objects in $\CM^D$. To this end one needs to first establish the
map between quantities on each side of this correspondence.
For unit mass, the quantities $\ell$ (see Eq.~\eqref{obeeladef}) and $j$
(see Eq.~\eqref{obejaot}) measure the (linear) size
of the horizon along the torus or rotation plane, respectively. Then
$a_H(\ell)$ for KK black holes on $\CM^{n+2} \times \T^2$ is analogous
(up to constants) to $a_H(j)$ for rotating black holes in $\CM^{n+4}$.

More precisely, although the normalization of magnitudes in
\eqref{obejaot} and \eqref{obeeladef} are different, the functional dependence
of $a_H$ on $\ell$ or $j$ must be parametrically the same in both
functions, at least in the regime where the analogy is precise.
As a check on this, note that the function $a_H(\ell)$ in \eqref{obeaHmem} for
the uniform black membrane exhibits exactly the same functional form
\eqref{obeaHrh} as $a_H(j)$ for the MP black hole in the ultra-spinning limit.
Similarly, \eqref{obeaHstr} for the localized black string shows
the same functional form as \eqref{obeaHrh} for the
black ring in the large $j$ limit.
The most important application of the analogy, though, is to non-uniform
membrane phases (see Eq.~\eqref{obeahnum}), providing information about the phases
of pinched rotating black holes and how they connect to MP black holes and black
rings.

\subsection{Phase diagram of neutral rotating black holes on $\CM^D$ \label{obesec:conj} }

We present here the main points of the
proposed phase diagram \cite{obeEmparan:2007wm}
of neutral rotating black holes (with one angular momentum) in asymptotically flat space
that follows from the analogy described above.
To this end, we recall that the phases of KK black holes on a
two-torus were discussed in Sec.~\ref{obesec:torp} and depicted in the
representative phase diagram Fig.~\ref{obefig:KKphases7}.

\runinhead{Main sequence.}
The analogy developed above suggests that the phase diagram of
rotating black holes in the range $j>j_{\rm mem}$ where MP black holes
behave like black membranes, is qualitatively the same as that for KK black holes
on the torus (see Fig.~\ref{obefig:KKphases7}), with a pinched (lumpy)
rotating black hole connecting the MP black hole with the black ring.
This phase is depicted  in Fig.~\ref{obefig:sevendphases} as a gray line
emerging tangentially from the MP black hole curve at a critical
value $j_{\rm GL}$ that is currently unknown. Arguments were given
in \cite{obeEmparan:2003sy} to the effect that $j_{\rm GL}\gsim j_{\rm
mem}$, consistent with the analogy. As one moves along the gray line in
Fig.~\ref{obefig:sevendphases} in the direction away from the MP curve,
the pinch at the rotation axis of these black holes grows deeper.
Eventually, as depicted in
Fig.~\ref{obefig:membranes}, the horizon pinches down to zero thickness at
the axis and then the solutions connect to the black ring phase.
Note also  that we may have the `swallowtail'
structure of first-order phase transitions (as depicted Fig.~\ref{obefig:sevendphases}),
or instead that of second-order phase transitions (see Fig.~4 of \cite{obeEmparan:2007wm}).
It may not be unreasonable to expect that a swallowtail appears
at least for the lowest dimensions $D=6,7,\dots$, since this is in
fact the same type of phase structure that appears for $D=5$.

\begin{figure}[ht]
\centerline{\includegraphics[width=11.8cm]{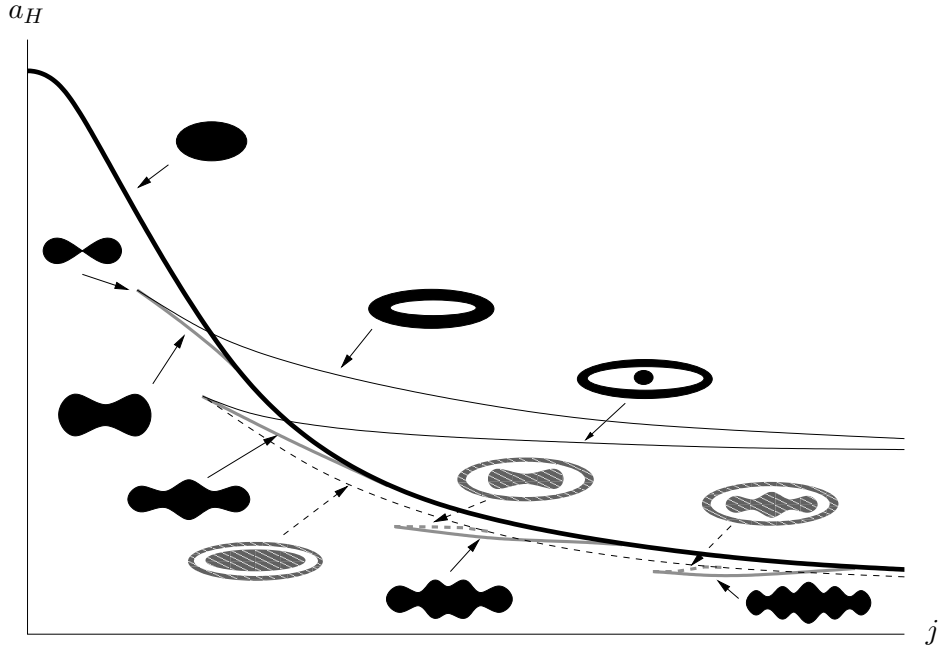}}
\begin{picture}(0,0)(0,0)
\put(40,250){$a_H$}
\put(387,15){$j$}
\end{picture}
\caption{\small Proposal for the phase diagram of thermal equilibrium phases
of rotating black holes in $D\geq 6$ with one angular momentum.
The solid lines and figures have significant arguments in their favor,
while the dashed lines and figures might not exist and
admit conceivable, but more complicated, alternatives. Some
features have been drawn arbitrarily, \eg
at any given bifurcation and in any dimension one may either have smooth connections
or swallowtails with cusps.  If thermal equilibrium is not imposed, the whole semi-infinite
strip $0<a_H <a_H(j=0)$, $0\leq j<\infty$ is covered, and multi-rings
are possible. Reprinted from Ref.~\cite{obeEmparan:2007wm}.}
\label{obefig:hidphases}
\end{figure}

Beyond this main sequence, Ref.~\cite{obeEmparan:2007wm} presents arguments
for further completion of the phase diagram, which is summarized in Fig.~\ref{obefig:hidphases}.
The most important features are as follows.

\runinhead{Infinite sequence of lumpy (pinched) black holes.}
Another observation based on the membrane analogy is that the
phase diagram of rotating black holes should also exhibit an
infinite sequence \cite{obeEmparan:2003sy,obeEmparan:2007wm}
of lumpy (pinched) black holes emerging from the curve of MP black holes at
increasing values of $j$. These are the analogues of the $k$-copied phases
in the phase diagram of KK black holes that appear at increasing
$\ell$ according to \eqref{obecopy}. In this connection note that for
the GL zero-modes of MP black holes one must choose axially-symmetric combinations,
implying a change of basis from plane waves $\exp(i k_{\rm GL} z)$ to Bessel
functions. Axially symmetric modes have a profile $J_0(k_{\rm
GL}a\sin\theta)$ \cite{obeEmparan:2003sy}. The main point here
is that the wavelength $\lambda_{\rm GL}$ (see \eqref{obelambgl}) of the GL
zero-mode remains the same in the two analogue systems, to first
approximation, even if the profiles are not the same. One is thus led
to the existence of an infinite sequence of
pinched black hole phases emanating from the MP curve at increasing
values $j_{\rm GL}^{(k)}$.

\runinhead{Black Saturn.}
If we focus on the first copy ($k=2$), on the KK black hole side this
corresponds to a non-uniform membrane on $\T^2$  with a GL zero-mode
perturbation of the membrane with two minima, which grows
to merge with a configuration of two identical black strings
localized on the torus. For the MP black hole, the analogue is the
development of a circular pinch, which then grows deeper until the
merger with a black Saturn configuration in thermal equilibrium. Thermal
equilibrium, \ie\ equal temperature and angular velocity on all
disconnected components of the event horizon, is in fact naturally
expected for solutions that merge with pinched black holes, since the
temperature and angular velocity of the latter should be uniform on the
horizon all the way down to the merger, and we do not expect them to
jump discontinuously there. These appear to be the natural higher-dimensional
generalization of the five-dimensional black Saturn \cite{obeElvang:2007rd},
and one may invoke the same arguments as those in Ref.~\cite{obeElvang:2007hg}.
When the size of the central black hole is small compared to the radius
of the black ring, the interaction between the two objects is
small and, to a first approximation, one can simply combine them
linearly. It follows that, under the assumption of equal
temperatures and angular velocities for the two black objects in the
black Saturn, as $j$ is increased a larger fraction of the total mass
and the total angular momentum is carried by the black ring, and less by
the central black hole. Then, this black Saturn curve must asymptote to
the curve of a single black ring.

\runinhead{Pancaked and pinched black Saturns.}
The existence of these phases and their appearance in the phase diagram
Fig.~\ref{obefig:hidphases} (in which they appear dashed) is based on comparatively
less compelling arguments. Nevertheless, these conjectural phases provide a simple
and natural way of completing the curves in the phase diagram that is
consistent with the available information. We refer to
Ref.~\cite{obeEmparan:2007wm} for further details on these phases.

It should also be noted that in the diagram of Fig.~\ref{obefig:hidphases}
only the thermal equilibrium phases among the possible multi-black hole phases
are represented. The existence of
multi-black rings, with or without a central black hole, in thermal
equilibrium is not expected. In general one does expect the
existence of multi-black ring configurations, possibly with a central
black hole, in which the different black objects have
different surface gravities and different angular velocities.
These configurations can be seen as the analogue of the multi-localized
string configurations on the torus that can be obtained from
multi-black hole configurations on the circle \cite{obeDias:2007hg}
discussed in Sec.~\ref{obesec:mubh} by adding a uniform direction.

\section{Outlook \label{obesec:outl}}

We conclude by briefly presenting a number of important issues and questions for
future research. See also the reviews
\cite{obeKol:2004ww,obeEmparan:2006mm,obeHarmark:2007md,obeEmparan:2008eg} for further
discussion and other open problems.

\runinhead{Stability.}
In both classes discussed in this lecture it would be interesting
to further study the stability of the various solutions. For KK black holes, this includes the classical stability of the non-uniform
black string and the localized black hole.
For the rotating black hole case, we note that
black rings at large $j$ in any $D\geq 5$ are expected
to suffer from a GL instability that creates ripples along the $S^1$
and presumably fragments the black ring into black holes flying
apart \cite{obeEmparan:2001wn,obeHovdebo:2006jy,obeElvang:2006dd}. This instability
may switch off at $j\sim O(1)$. In analogy to the five-dimensional
case \cite{obeArcioni:2004ww,obeElvang:2006dd}, one could also study turning points of $j$.
If these are absent, pinched black holes would presumably be stable to radial perturbations.

\runinhead{Other compactified solutions.}
It would also be interesting to examine the existence of other
classes of solutions with a compactified direction. For example in
Ref.~\cite{obeMaeda:2006hd} a supersymmetric rotating black hole in a
compactified space-time was found and charged black holes in
compactified space-times are considered in Ref.~\cite{obeKarlovini:2005cn}.
In another direction, new solutions with Kaluza-Klein boundary conditions
for anti-de-Sitter spacetimes have recently been constructed in
Refs.~\cite{obeCopsey:2006br,obeMann:2006yi}. Finally, rotating non-uniform solutions
in KK space have been constructed numerically in Ref.~\cite{obeKleihaus:2007dg}
(see also Ref.~\cite{obeKleihaus:2007kc}).

\runinhead{Numerical solutions.}
For both classes of higher-dimensional black holes presented in this lecture,
it would be interesting to attempt to further apply numerical techniques
in order to construct the new solutions. For example, for
multi-black hole configurations on the cylinder this could confirm whether
there are multi-black hole solutions for which the temperatures
converge when approaching the merger points (as discussed in Ref.~\cite{obeDias:2007hg}).
Furthermore, one could try to confirm the existence of the conjectured lumpy black
holes (see Sec.~\ref{obesec:copd}). Similarly, for rotating black holes, numerical
construction of the entire black ring phase and of the pinched black hole phase
would be very interesting.

\runinhead{Effective field theory techniques.}
As mentioned in Sec.~\ref{obesec:solm}, an alternative to the matched
expansion is the use of classical effective field theory \cite{obeChu:2006ce,obeKol:2007rx}
to obtain the corrected thermodynamics of new solutions in a perturbative expansion.
It would be interesting to use this method to go beyond the first order for
the solutions discussed in this lecture and apply it to other
 extended brane-like black holes with or without rotation.

\runinhead{Other black rings.}
The method used to construct thin black rings in asymptotically flat space
can also be used to study thin black
rings in external gravitational potentials, yielding \eg black Saturn
or black rings in AdS or dS spacetime.%
\footnote{In \cite{obeKunduri:2006uh} the existence of supersymmetric
black rings in AdS is considered.}
Similarly, one could study black rings with charges
\cite{obeElvang:2003yy,obeElvang:2004rt} and with dipoles
\cite{obeEmparan:2004wy}. In this connection we note that the existence
of small supersymmetric black rings in $D\geq 5$ was argued in
\cite{obeDabholkar:2006za}.

\runinhead{More rotation parameters.}
One may try to extended the analysis to  black rings with horizon $S^1\times
S^{n+1}$ with rotation not only along $S^1$ but also in the $S^{n+1}$.
Rotation in the $S^{n+1}$ will introduce particularly rich dynamics for $n\geq 3$,
 since it is then possible to have ultra-spinning regimes for this rotation too,
 leading to pinches of the $S^{n+1}$ and further connections to phases with
horizon $S^1\times S^1\times S^{n}$, and so forth.

\runinhead{Blackfolds.}
Following the construction in Sec.~\ref{obesec:robh}, one can envision many
generalizations. In this way one could study the possible existence of more general
blackfolds, obtained by
taking a black $p$-brane with horizon topology $\R^{p} \times
S^{q}$ and bending $\R^{p}$ to form some compact manifold. One must
then find out under which conditions a curved black $p$-brane can
satisfy the equilibrium equation \eqref{obeKT}. This method is constructive
and uses dynamical information to determine possible horizon geometries.
In contrast, conventional approaches
based on topological considerations are non-constructive and
have only found very weak restrictions in six or more dimensions
\cite{obeHelfgott:2005jn,obeGalloway:2005mf}.

\runinhead{Plasma balls and rings.}
There is also another more indirect approach to higher-dimensional black rings in AdS,
using the AdS/CFT correspondence. In Ref.~\cite{obeLahiri:2007ae}
stationary, axially symmetric spinning configurations of plasma
in ${\cal N}=4$ SYM theory compactified to $d=3$ on a Scherk-Schwarz
circle were studied. On the gravity side, these correspond
to large rotating black holes and black rings in the dual Scherk-Schwarz
compactified AdS$_5$ space. Interestingly, the phase diagram of these
rotating fluid configurations, even if dual to black holes larger than
the AdS radius, reproduces many of the qualitative features of the MP
black holes and black rings in five-dimensional flat spacetime.
Higher-dimensional generalizations of this setup give predictions for
the phases of black holes in Scherk-Schwarz compactified AdS$_D$ with
$D>5$. In this way, evidence was found \cite{obeLahiri:2007ae}
for rotating black rings and `pinched' black holes in AdS$_6$,
that can be considered the AdS-analogues of the phases conjectured in
\cite{obeEmparan:2003sy,obeEmparan:2007wm}, discussed in Secs.~\ref{obesec:robh}
and \ref{obesec:phas}.

\runinhead{Microscopic entropy for three-charge black holes.}
One could extend the work of Ref.~\cite{obeHarmark:2006df,obeHarmark:2007uy}
by applying the boost/U-duality map of \cite{obeHarmark:2004ws} to the
multi-black hole configurations of Ref.~\cite{obeDias:2007hg}.
In particular, this would enable to compute the  first correction to
the finite entropy of the resulting three-charge multi-black hole configurations
on a circle. It would be interesting to then try to derive these expressions from
a microscopic calculation following the single three-charge black hole
case considered in Refs.~\cite{obeHarmark:2006df,obeChowdhury:2006qn}.

\runinhead{Braneworld black holes.}
The higher dimensional black holes and branes described in this lecture
also appear naturally in the discussion of the braneworld model
of large extra dimensions \cite{obeArkani-Hamed:1998rs,obeAntoniadis:1998ig}. In other braneworld models such as the one
proposed by Randall and Sundrum \cite{obeRandall:1999ee,obeRandall:1999vf} the geometry is warped in the extra direction and
the discovery of black hole solutions in this context has proven
more difficult. It would be interesting to consider the higher-dimensional
black hole solutions considered in this lecture in these contexts.

\section*{Acknowledgements}

I would like to thank the organizers, especially Lefteris Papantonopoulos,
 of the Fourth Aegean Summer School on
Black Holes (Sept. 17-22, 2007, Mytiline, Island of Lesvos, Greece)
 for a stimulating and interesting school. I also thank
 Oscar Dias,  Roberto Emparan, Troels Harmark, Rob Myers, Vasilis Niarchos
 and  Maria Jose Rodriguez for collaboration on the work presented here.
This work is partially supported by the European Community's Human
Potential Programme under contract MRTN-CT-2004-005104
`Constituents, fundamental forces and symmetries of the universe'.


\begin{thebibliography}{10%
0}

\addcontentsline{toc}{section}{References}

\bibitem{obeKol:2004ww}
B.~Kol, ``The phase transition between caged black holes and black strings: {A}
  review,'' {\em Phys. Rept.} {\bf 422} (2006) 119--165,
\href{http://arXiv.org/abs/hep-th/0411240}{{\tt hep-th/0411240}}.

\bibitem{obeEmparan:2006mm}
R.~Emparan and H.~S. Reall, ``Black rings,'' {\em Class. Quant. Grav.} {\bf 23}
  (2006) R169,
\href{http://arXiv.org/abs/hep-th/0608012}{{\tt hep-th/0608012}}.

\bibitem{obeHarmark:2007md}
T.~Harmark, V.~Niarchos, and N.~A. Obers, ``Instabilities of black strings and
  branes,'' {\em Class. Quant. Grav.} {\bf 24} (2007) R1--R90,
\href{http://arXiv.org/abs/hep-th/0701022}{{\tt hep-th/0701022}}.

\bibitem{obeEmparan:2008eg}
R.~Emparan and H.~S. Reall, ``{Black Holes in Higher Dimensions},''
\href{http://arXiv.org/abs/arXiv:0801.3471 [hep-th]}{{\tt arXiv:0801.3471
  [hep-th]}}.

\bibitem{obeStrominger:1996sh}
A.~Strominger and C.~Vafa, ``Microscopic origin of the {Bekenstein-Hawking}
  entropy,'' {\em Phys. Lett.} {\bf B379} (1996) 99--104,
\href{http://arXiv.org/abs/hep-th/9601029}{{\tt hep-th/9601029}}.

\bibitem{obeMathur:2005zp}
S.~D. Mathur, ``The fuzzball proposal for black holes: {A}n elementary
  review,'' {\em Fortsch. Phys.} {\bf 53} (2005) 793--827,
\href{http://arXiv.org/abs/hep-th/0502050}{{\tt hep-th/0502050}}.

\bibitem{obeMathur:2005ai}
S.~D. Mathur, ``The quantum structure of black holes,'' {\em Class. Quant.
  Grav.} {\bf 23} (2006) R115,
\href{http://arXiv.org/abs/hep-th/0510180}{{\tt hep-th/0510180}}.

\bibitem{obeMaldacena:1997re}
J.~M. Maldacena, ``The large {$N$} limit of superconformal field theories and
  supergravity,'' {\em Adv. Theor. Math. Phys.} {\bf 2} (1998) 231--252,
\href{http://arXiv.org/abs/hep-th/9711200}{{\tt hep-th/9711200}}.

\bibitem{obeAharony:1999ti}
O.~Aharony, S.~S. Gubser, J.~Maldacena, H.~Ooguri, and Y.~Oz, ``Large {$N$}
  field theories, string theory and gravity,'' {\em Phys. Rept.} {\bf 323}
  (2000) 183,
\href{http://arXiv.org/abs/hep-th/9905111}{{\tt hep-th/9905111}}.

\bibitem{obeAharony:2004ig}
O.~Aharony, J.~Marsano, S.~Minwalla, and T.~Wiseman, ``Black hole - black
  string phase transitions in thermal 1+1 dimensional supersymmetric
  {Yang-Mills} theory on a circle,'' {\em Class. Quant. Grav.} {\bf 21} (2004)
  5169--5192,
\href{http://arXiv.org/abs/hep-th/0406210}{{\tt hep-th/0406210}}.

\bibitem{obeHarmark:2004ws}
T.~Harmark and N.~A. Obers, ``New phases of near-extremal branes on a circle,''
  {\em JHEP} {\bf 09} (2004) 022,
\href{http://arXiv.org/abs/hep-th/0407094}{{\tt hep-th/0407094}}.

\bibitem{obeArkaniHamed:1998rs}
N.~Arkani-Hamed, S.~Dimopoulos, and G.~R. Dvali, ``The hierarchy problem and
  new dimensions at a millimeter,'' {\em Phys. Lett.} {\bf B429} (1998)
  263--272,
\href{http://arXiv.org/abs/hep-ph/9803315}{{\tt hep-ph/9803315}}.

\bibitem{obeAntoniadis:1998ig}
I.~Antoniadis, N.~Arkani-Hamed, S.~Dimopoulos, and G.~R. Dvali, ``New
  dimensions at a millimeter to a {F}ermi and superstrings at a {TeV},'' {\em
  Phys. Lett.} {\bf B436} (1998) 257--263,
\href{http://arXiv.org/abs/hep-ph/9804398}{{\tt hep-ph/9804398}}.

\bibitem{obeKanti:2004nr}
P.~Kanti, ``Black holes in theories with large extra dimensions: {A} review,''
  {\em Int. J. Mod. Phys.} {\bf A19} (2004) 4899--4951,
\href{http://arXiv.org/abs/hep-ph/0402168}{{\tt hep-ph/0402168}}.

\bibitem{obeMyers:1986un}
R.~C. Myers and M.~J. Perry, ``Black holes in higher dimensional space-times,''
  {\em Ann. Phys.} {\bf 172} (1986)
304.

\bibitem{obeEmparan:2001wn}
R.~Emparan and H.~S. Reall, ``A rotating black ring in five dimensions,'' {\em
  Phys. Rev. Lett.} {\bf 88} (2002) 101101,
\href{http://arXiv.org/abs/hep-th/0110260}{{\tt hep-th/0110260}}.

\bibitem{obeElvang:2007rd}
H.~Elvang and P.~Figueras, ``Black saturn,'' {\em JHEP} {\bf 05} (2007) 050,
\href{http://arXiv.org/abs/hep-th/0701035}{{\tt hep-th/0701035}}.

\bibitem{obeElvang:2007hg}
H.~Elvang, R.~Emparan, and P.~Figueras, ``Phases of five-dimensional black
  holes,'' {\em JHEP} {\bf 05} (2007) 056,
\href{http://arXiv.org/abs/hep-th/0702111}{{\tt hep-th/0702111}}.

\bibitem{obeIguchi:2007is}
H.~Iguchi and T.~Mishima, ``Black di-ring and infinite nonuniqueness,'' {\em
  Phys. Rev.} {\bf D75} (2007) 064018,
\href{http://arXiv.org/abs/hep-th/0701043}{{\tt hep-th/0701043}}.

\bibitem{obeEvslin:2007fv}
J.~Evslin and C.~Krishnan, ``The black di-ring: An inverse scattering
  construction,''
\href{http://arXiv.org/abs/arXiv:0706.1231 [hep-th]}{{\tt arXiv:0706.1231
  [hep-th]}}.

\bibitem{obePomeransky:2006bd}
A.~A. Pomeransky and R.~A. Sen'kov, ``Black ring with two angular momenta,''
\href{http://arXiv.org/abs/hep-th/0612005}{{\tt hep-th/0612005}}.

\bibitem{obeIzumi:2007qx}
K.~Izumi, ``Orthogonal black di-ring solution,''
\href{http://arXiv.org/abs/arXiv:0712.0902 [hep-th]}{{\tt arXiv:0712.0902
  [hep-th]}}.

\bibitem{obeElvang:2007hs}
H.~Elvang and M.~J. Rodriguez, ``Bicycling black rings,''
\href{http://arXiv.org/abs/arXiv:0712.2425 [hep-th]}{{\tt arXiv:0712.2425
  [hep-th]}}.

\bibitem{obeEmparan:2001wk}
R.~Emparan and H.~S. Reall, ``Generalized {Weyl} solutions,'' {\em Phys. Rev.}
  {\bf D65} (2002) 084025,
\href{http://arXiv.org/abs/hep-th/0110258}{{\tt hep-th/0110258}}.

\bibitem{obeHarmark:2004rm}
T.~Harmark, ``Stationary and axisymmetric solutions of higher-dimensional
  general relativity,'' {\em Phys. Rev.} {\bf D70} (2004) 124002,
\href{http://arXiv.org/abs/hep-th/0408141}{{\tt hep-th/0408141}}.

\bibitem{obeBelinsky:1971nt}
V.~A. Belinsky and V.~E. Zakharov, ``Integration of the {Einstein} equations by
  the inverse scattering problem technique and the calculation of the exact
  soliton solutions,'' {\em Sov. Phys. JETP} {\bf 48} (1978)
985--994.

\bibitem{obeBelinsky:1979}
V.~A. Belinsky and V.~E. Zakharov, ``Stationary gravitational solitons with
  axial symmetry,'' {\em Sov. Phys. JETP} {\bf 50} (1979) 1.

\bibitem{obeBelinski:2001ph}
V.~Belinski and E.~Verdaguer, ``Gravitational solitons,''. Cambridge, UK: Univ.
  Pr. (2001) 258 p.

\bibitem{obePomeransky:2005sj}
A.~A. Pomeransky, ``Complete integrability of higher-dimensional {E}instein
  equations with additional symmetry, and rotating black holes,'' {\em Phys.
  Rev.} {\bf D73} (2006) 044004,
\href{http://arXiv.org/abs/hep-th/0507250}{{\tt hep-th/0507250}}.

\bibitem{obeEmparan:2007wm}
R.~Emparan, T.~Harmark, V.~Niarchos, N.~A. Obers, and M.~J. Rodriguez, ``The
  phase structure of higher-dimensional black rings and black holes,'' {\em
  JHEP} {\bf 10} (2007) 110,
\href{http://arXiv.org/abs/arXiv:0708.2181 [hep-th]}{{\tt arXiv:0708.2181
  [hep-th]}}.

\bibitem{obeHarmark:2003yz}
T.~Harmark, ``Small black holes on cylinders,'' {\em Phys. Rev.} {\bf D69}
  (2004) 104015,
\href{http://arXiv.org/abs/hep-th/0310259}{{\tt hep-th/0310259}}.

\bibitem{obeGorbonos:2004uc}
D.~Gorbonos and B.~Kol, ``A dialogue of multipoles: Matched asymptotic
  expansion for caged black holes,'' {\em JHEP} {\bf 06} (2004) 053,
\href{http://arXiv.org/abs/hep-th/0406002}{{\tt hep-th/0406002}}.

\bibitem{obeKarasik:2004ds}
D.~Karasik, C.~Sahabandu, P.~Suranyi, and L.~C.~R. Wijewardhana, ``Analytic
  approximation to 5 dimensional black holes with one compact dimension,'' {\em
  Phys. Rev.} {\bf D71} (2005) 024024,
\href{http://arXiv.org/abs/hep-th/0410078}{{\tt hep-th/0410078}}.

\bibitem{obeGorbonos:2005px}
D.~Gorbonos and B.~Kol, ``Matched asymptotic expansion for caged black holes:
  Regularization of the post-{N}ewtonian order,'' {\em Class. Quant. Grav.}
  {\bf 22} (2005) 3935--3960,
\href{http://arXiv.org/abs/hep-th/0505009}{{\tt hep-th/0505009}}.

\bibitem{obeDias:2007hg}
O.~J.~C. Dias, T.~Harmark, R.~C. Myers, and N.~A. Obers, ``Multi-black hole
  configurations on the cylinder,'' {\em Phys. Rev.} {\bf D76} (2007) 104025,
\href{http://arXiv.org/abs/arXiv:0706.3645 [hep-th]}{{\tt arXiv:0706.3645
  [hep-th]}}.

\bibitem{obeEmparan:2003sy}
R.~Emparan and R.~C. Myers, ``Instability of ultra-spinning black holes,'' {\em
  JHEP} {\bf 09} (2003) 025,
\href{http://arXiv.org/abs/hep-th/0308056}{{\tt hep-th/0308056}}.

\bibitem{obeGubser:2001ac}
S.~S. Gubser, ``On non-uniform black branes,'' {\em Class. Quant. Grav.} {\bf
  19} (2002) 4825--4844,
\href{http://arXiv.org/abs/hep-th/0110193}{{\tt hep-th/0110193}}.

\bibitem{obeWiseman:2002zc}
T.~Wiseman, ``Static axisymmetric vacuum solutions and non-uniform black
  strings,'' {\em Class. Quant. Grav.} {\bf 20} (2003) 1137--1176,
\href{http://arXiv.org/abs/hep-th/0209051}{{\tt hep-th/0209051}}.

\bibitem{obeSorkin:2004qq}
E.~Sorkin, ``A critical dimension in the black-string phase transition,'' {\em
  Phys. Rev. Lett.} {\bf 93} (2004) 031601,
\href{http://arXiv.org/abs/hep-th/0402216}{{\tt hep-th/0402216}}.

\bibitem{obeKleihaus:2006ee}
B.~Kleihaus, J.~Kunz, and E.~Radu, ``New nonuniform black string solutions,''
  {\em JHEP} {\bf 06} (2006) 016,
\href{http://arXiv.org/abs/hep-th/0603119}{{\tt hep-th/0603119}}.

\bibitem{obeSorkin:2006wp}
E.~Sorkin, ``Non-uniform black strings in various dimensions,'' {\em Phys.
  Rev.} {\bf D74} (2006) 104027,
\href{http://arXiv.org/abs/gr-qc/0608115}{{\tt gr-qc/0608115}}.

\bibitem{obeKleihaus:2007cf}
B.~Kleihaus and J.~Kunz, ``Interior of nonuniform black strings,''
\href{http://arXiv.org/abs/arXiv:0710.1726 [hep-th]}{{\tt arXiv:0710.1726
  [hep-th]}}.

\bibitem{obeHarmark:2002tr}
T.~Harmark and N.~A. Obers, ``Black holes on cylinders,'' {\em JHEP} {\bf 05}
  (2002) 032,
\href{http://arXiv.org/abs/hep-th/0204047}{{\tt hep-th/0204047}}.

\bibitem{obeChu:2006ce}
Y.-Z. Chu, W.~D. Goldberger, and I.~Z. Rothstein, ``Asymptotics of
  {$d$}-dimensional {Kaluza-Klein} black holes: Beyond the {N}ewtonian
  approximation,'' {\em JHEP} {\bf 03} (2006) 013,
\href{http://arXiv.org/abs/hep-th/0602016}{{\tt hep-th/0602016}}.

\bibitem{obeKol:2007rx}
B.~Kol and M.~Smolkin, ``Classical effective field theory and caged black
  holes,''
\href{http://arXiv.org/abs/arXiv:0712.2822 [hep-th]}{{\tt arXiv:0712.2822
  [hep-th]}}.

\bibitem{obeSorkin:2003ka}
E.~Sorkin, B.~Kol, and T.~Piran, ``Caged black holes: Black holes in
  compactified spacetimes. {II}: 5d numerical implementation,'' {\em Phys.
  Rev.} {\bf D69} (2004) 064032,
\href{http://arXiv.org/abs/hep-th/0310096}{{\tt hep-th/0310096}}.

\bibitem{obeKudoh:2003ki}
H.~Kudoh and T.~Wiseman, ``Properties of {Kaluza-Klein} black holes,'' {\em
  Prog. Theor. Phys.} {\bf 111} (2004) 475--507,
\href{http://arXiv.org/abs/hep-th/0310104}{{\tt hep-th/0310104}}.

\bibitem{obeKudoh:2004hs}
H.~Kudoh and T.~Wiseman, ``Connecting black holes and black strings,'' {\em
  Phys. Rev. Lett.} {\bf 94} (2005) 161102,
\href{http://arXiv.org/abs/hep-th/0409111}{{\tt hep-th/0409111}}.

\bibitem{obeElvang:2004iz}
H.~Elvang, T.~Harmark, and N.~A. Obers, ``Sequences of bubbles and holes: New
  phases of {Kaluza-Klein} black holes,'' {\em JHEP} {\bf 01} (2005) 003,
\href{http://arXiv.org/abs/hep-th/0407050}{{\tt hep-th/0407050}}.

\bibitem{obeHarmark:2003dg}
T.~Harmark and N.~A. Obers, ``New phase diagram for black holes and strings on
  cylinders,'' {\em Class. Quantum Grav.} {\bf 21} (2004) 1709--1724,
\href{http://arXiv.org/abs/hep-th/0309116}{{\tt hep-th/0309116}}.

\bibitem{obeKol:2003if}
B.~Kol, E.~Sorkin, and T.~Piran, ``Caged black holes: Black holes in
  compactified spacetimes. {I}: Theory,'' {\em Phys. Rev.} {\bf D69} (2004)
  064031,
\href{http://arXiv.org/abs/hep-th/0309190}{{\tt hep-th/0309190}}.

\bibitem{obeHarmark:2003eg}
T.~Harmark and N.~A. Obers, ``Phase structure of black holes and strings on
  cylinders,'' {\em Nucl. Phys.} {\bf B684} (2004) 183--208,
\href{http://arXiv.org/abs/hep-th/0309230}{{\tt hep-th/0309230}}.

\bibitem{obeGregory:1993vy}
R.~Gregory and R.~Laflamme, ``Black strings and {$p$}-branes are unstable,''
  {\em Phys. Rev. Lett.} {\bf 70} (1993) 2837--2840,
\href{http://arXiv.org/abs/hep-th/9301052}{{\tt hep-th/9301052}}.

\bibitem{obeGregory:1994bj}
R.~Gregory and R.~Laflamme, ``The instability of charged black strings and
  {$p$}-branes,'' {\em Nucl. Phys.} {\bf B428} (1994) 399--434,
\href{http://arXiv.org/abs/hep-th/9404071}{{\tt hep-th/9404071}}.

\bibitem{obeKol:2002xz}
B.~Kol, ``Topology change in general relativity and the black-hole black-string
  transition,''
\href{http://arXiv.org/abs/hep-th/0206220}{{\tt hep-th/0206220}}.

\bibitem{obeWiseman:2002ti}
T.~Wiseman, ``From black strings to black holes,'' {\em Class. Quant. Grav.}
  {\bf 20} (2003) 1177--1186,
\href{http://arXiv.org/abs/hep-th/0211028}{{\tt hep-th/0211028}}.

\bibitem{obeKol:2003ja}
B.~Kol and T.~Wiseman, ``Evidence that highly non-uniform black strings have a
  conical waist,'' {\em Class. Quant. Grav.} {\bf 20} (2003) 3493--3504,
\href{http://arXiv.org/abs/hep-th/0304070}{{\tt hep-th/0304070}}.

\bibitem{obeKol:2004pn}
B.~Kol and E.~Sorkin, ``On black-brane instability in an arbitrary dimension,''
  {\em Class. Quant. Grav.} {\bf 21} (2004) 4793--4804,
\href{http://arXiv.org/abs/gr-qc/0407058}{{\tt gr-qc/0407058}}.

\bibitem{obeKol:2006vu}
B.~Kol and E.~Sorkin, ``{LG} ({Landau-Ginzburg}) in {GL}
  ({Gregory-Laflamme}),'' {\em Class. Quant. Grav.} {\bf 23} (2006) 4563--4592,
\href{http://arXiv.org/abs/hep-th/0604015}{{\tt hep-th/0604015}}.

\bibitem{obeIsrael:1967wq}
W.~Israel, ``Event horizons in static vacuum space-times,'' {\em Phys. Rev.}
  {\bf 164} (1967)
1776--1779.

\bibitem{obeCarter:1971}
B.~Carter, ``Axisymmetric black hole has only two degrees of freedom,'' {\em
  Phys. Rev. Lett.} {\bf 26} (1971) 331--333.

\bibitem{obeHawking:1972vc}
S.~W. Hawking, ``Black holes in {General Relativity},'' {\em Commun. Math.
  Phys.} {\bf 25} (1972)
152--166.

\bibitem{obeRobinson:1975}
D.~C. Robinson, ``Uniqueness of the {Kerr} black hole,'' {\em Phys. Rev. Lett.}
  {\bf 34} (1975) 905--906.

\bibitem{obeRegge:1957td}
T.~Regge, ``Stability of a Schwarzschild singularity,'' {\em Phys. Rev.} {\bf
  108} (1957)
1063--1069.

\bibitem{obeZerilli:1971wd}
F.~J. Zerilli, ``Gravitational field of a particle falling in a {S}chwarzschild
  geometry analyzed in tensor harmonics,'' {\em Phys. Rev.} {\bf D2} (1970)
2141--2160.

\bibitem{obeTeukolsky:1973ha}
S.~A. Teukolsky, ``Perturbations of a rotating black hole. 1. {F}undamental
  equations for gravitational electromagnetic, and neutrino field
  perturbations,'' {\em Astrophys. J.} {\bf 185} (1973)
635--647.

\bibitem{obeKodama:2007ph}
H.~Kodama, ``Perturbations and stability of higher-dimensional black holes,''
\href{http://arXiv.org/abs/arXiv:0712.2703 [hep-th]}{{\tt arXiv:0712.2703
  [hep-th]}}.

\bibitem{obeTangherlini:1963}
F.~R. Tangherlini, ``Schwarzschild field in {\sl n} dimensions and the
  dimensionality of space problem,'' {\em Nuovo Cimento} {\bf 27} (1963) 636.

\bibitem{obeGibbons:2002bh}
G.~W. Gibbons, D.~Ida, and T.~Shiromizu, ``Uniqueness and non-uniqueness of
  static vacuum black holes in higher dimensions,'' {\em Prog. Theor. Phys.
  Suppl.} {\bf 148} (2003) 284--290,
\href{http://arXiv.org/abs/gr-qc/0203004}{{\tt gr-qc/0203004}}.

\bibitem{obeGibbons:2002av}
G.~W. Gibbons, D.~Ida, and T.~Shiromizu, ``Uniqueness and non-uniqueness of
  static black holes in higher dimensions,'' {\em Phys. Rev. Lett.} {\bf 89}
  (2002) 041101,
\href{http://arXiv.org/abs/hep-th/0206049}{{\tt hep-th/0206049}}.

\bibitem{obeKodama:2003jz}
H.~Kodama and A.~Ishibashi, ``A master equation for gravitational perturbations
  of maximally symmetric black holes in higher dimensions,'' {\em Prog. Theor.
  Phys.} {\bf 110} (2003) 701--722,
\href{http://arXiv.org/abs/hep-th/0305147}{{\tt hep-th/0305147}}.

\bibitem{obeIshibashi:2003ap}
A.~Ishibashi and H.~Kodama, ``Stability of higher-dimensional {S}chwarzschild
  black holes,'' {\em Prog. Theor. Phys.} {\bf 110} (2003) 901--919,
\href{http://arXiv.org/abs/hep-th/0305185}{{\tt hep-th/0305185}}.

\bibitem{obeKodama:2003kk}
H.~Kodama and A.~Ishibashi, ``Master equations for perturbations of generalized
  static black holes with charge in higher dimensions,'' {\em Prog. Theor.
  Phys.} {\bf 111} (2004) 29--73,
\href{http://arXiv.org/abs/hep-th/0308128}{{\tt hep-th/0308128}}.

\bibitem{obeHollands:2006rj}
S.~Hollands, A.~Ishibashi, and R.~M. Wald, ``{A higher dimensional stationary
  rotating black hole must be axisymmetric},'' {\em Commun. Math. Phys.} {\bf
  271} (2007) 699--722,
\href{http://arXiv.org/abs/gr-qc/0605106}{{\tt gr-qc/0605106}}.

\bibitem{obeMorisawa:2004tc}
Y.~Morisawa and D.~Ida, ``A boundary value problem for the five-dimensional
  stationary rotating black holes,'' {\em Phys. Rev.} {\bf D69} (2004) 124005,
\href{http://arXiv.org/abs/gr-qc/0401100}{{\tt gr-qc/0401100}}.

\bibitem{obeHollands:2007aj}
S.~Hollands and S.~Yazadjiev, ``Uniqueness theorem for 5-dimensional black
  holes with two axial Killing fields,''
\href{http://arXiv.org/abs/arXiv:0707.2775 [gr-qc]}{{\tt arXiv:0707.2775
  [gr-qc]}}.

\bibitem{obeGiusto:2007fx}
S.~Giusto and A.~Saxena, ``{Stationary axisymmetric solutions of five
  dimensional gravity},'' {\em Class. Quant. Grav.} {\bf 24} (2007) 4269--4294,
\href{http://arXiv.org/abs/arXiv:0705.4484 [hep-th]}{{\tt arXiv:0705.4484
  [hep-th]}}.

\bibitem{obeGoldberger:2004jt}
W.~D. Goldberger and I.~Z. Rothstein, ``An effective field theory of gravity
  for extended objects,'' {\em Phys. Rev.} {\bf D73} (2006) 104029,
\href{http://arXiv.org/abs/hep-th/0409156}{{\tt hep-th/0409156}}.

\bibitem{obeHarmark:2005pp}
T.~Harmark and N.~A. Obers, ``Phases of {Kaluza-Klein} black holes: {A} brief
  review,''
\href{http://arXiv.org/abs/hep-th/0503020}{{\tt hep-th/0503020}}.

\bibitem{obeHarmark:2004ch}
T.~Harmark and N.~A. Obers, ``General definition of gravitational tension,''
  {\em JHEP} {\bf 05} (2004) 043,
\href{http://arXiv.org/abs/hep-th/0403103}{{\tt hep-th/0403103}}.

\bibitem{obeMyers:1999ps}
R.~C. Myers, ``Stress tensors and {Casimir} energies in the {AdS/CFT}
  correspondence,'' {\em Phys. Rev.} {\bf D60} (1999) 046002,
\href{http://arXiv.org/abs/hep-th/9903203}{{\tt hep-th/9903203}}.

\bibitem{obeTraschen:2001pb}
J.~H. Traschen and D.~Fox, ``Tension perturbations of black brane spacetimes,''
  {\em Class. Quant. Grav.} {\bf 21} (2004) 289--306,
\href{http://arXiv.org/abs/gr-qc/0103106}{{\tt gr-qc/0103106}}.

\bibitem{obeTownsend:2001rg}
P.~K. Townsend and M.~Zamaklar, ``The first law of black brane mechanics,''
  {\em Class. Quant. Grav.} {\bf 18} (2001) 5269--5286,
\href{http://arXiv.org/abs/hep-th/0107228}{{\tt hep-th/0107228}}.

\bibitem{obeKastor:2006ti}
D.~Kastor and J.~Traschen, ``Stresses and strains in the first law for
  {Kaluza-Klein} black holes,'' {\em JHEP} {\bf 09} (2006) 022,
\href{http://arXiv.org/abs/hep-th/0607051}{{\tt hep-th/0607051}}.

\bibitem{obeTraschen:2003jm}
J.~H. Traschen, ``A positivity theorem for gravitational tension in brane
  spacetimes,'' {\em Class. Quant. Grav.} {\bf 21} (2004) 1343--1350,
\href{http://arXiv.org/abs/hep-th/0308173}{{\tt hep-th/0308173}}.

\bibitem{obeShiromizu:2003gc}
T.~Shiromizu, D.~Ida, and S.~Tomizawa, ``Kinematical bound in asymptotically
  translationally invariant spacetimes,'' {\em Phys. Rev.} {\bf D69} (2004)
  027503,
\href{http://arXiv.org/abs/gr-qc/0309061}{{\tt gr-qc/0309061}}.

\bibitem{obeHarmark:2003fz}
T.~Harmark and N.~A. Obers, ``Black holes and black strings on cylinders,''
  {\em Fortsch. Phys.} {\bf 51} (2003) 793--798,
\href{http://arXiv.org/abs/hep-th/0301020}{{\tt hep-th/0301020}}.

\bibitem{obeKol:2006ga}
B.~Kol, ``The power of action: {'The'} derivation of the black hole negative
  mode,''
\href{http://arXiv.org/abs/hep-th/0608001}{{\tt hep-th/0608001}}.

\bibitem{obeKol:2006ux}
B.~Kol, ``Perturbations around backgrounds with one non-homogeneous
  dimension,''
\href{http://arXiv.org/abs/hep-th/0609001}{{\tt hep-th/0609001}}.

\bibitem{obeCardoso:2006ks}
V.~Cardoso and O.~J.~C. Dias, ``{Rayleigh-Plateau} and {Gregory-Laflamme}
  instabilities of black strings,'' {\em Phys. Rev. Lett.} {\bf 96} (2006)
  181601,
\href{http://arXiv.org/abs/hep-th/0602017}{{\tt hep-th/0602017}}.

\bibitem{obeCardoso:2006sj}
V.~Cardoso and L.~Gualtieri, ``Equilibrium configurations of fluids and their
  stability in higher dimensions,'' {\em Class. Quant. Grav.} {\bf 23} (2006)
  7151--7198,
\href{http://arXiv.org/abs/hep-th/0610004}{{\tt hep-th/0610004}}.

\bibitem{obeGregory:1988nb}
R.~Gregory and R.~Laflamme, ``Hypercylindrical black holes,'' {\em Phys. Rev.}
  {\bf D37} (1988)
305.

\bibitem{obeMyers:1987rx}
R.~C. Myers, ``Higher dimensional black holes in compactified space- times,''
  {\em Phys. Rev.} {\bf D35} (1987)
455.

\bibitem{obeBogojevic:1991hv}
A.~R. Bogojevic and L.~Perivolaropoulos, ``Black holes in a periodic
  universe,'' {\em Mod. Phys. Lett.} {\bf A6} (1991)
369--376.

\bibitem{obeKorotkin:1994dw}
D.~Korotkin and H.~Nicolai, ``A periodic analog of the {Schwarzschild}
  solution,''
\href{http://arXiv.org/abs/gr-qc/9403029}{{\tt gr-qc/9403029}}.

\bibitem{obeFrolov:2003kd}
A.~V. Frolov and V.~P. Frolov, ``Black holes in a compactified spacetime,''
  {\em Phys. Rev.} {\bf D67} (2003) 124025,
\href{http://arXiv.org/abs/hep-th/0302085}{{\tt hep-th/0302085}}.

\bibitem{obeDeSmet:2002fv}
P.-J. De~Smet, ``Black holes on cylinders are not algebraically special,'' {\em
  Class. Quant. Grav.} {\bf 19} (2002) 4877--4896,
\href{http://arXiv.org/abs/hep-th/0206106}{{\tt hep-th/0206106}}.

\bibitem{obeHorowitz:2002dc}
G.~T. Horowitz, ``Playing with black strings,''
\href{http://arXiv.org/abs/hep-th/0205069}{{\tt hep-th/0205069}}.

\bibitem{obeEmparan:2004wy}
R.~Emparan, ``Rotating circular strings, and infinite non-uniqueness of black
  rings,'' {\em JHEP} {\bf 03} (2004) 064,
\href{http://arXiv.org/abs/hep-th/0402149}{{\tt hep-th/0402149}}.

\bibitem{obeCardoso:2007ka}
V.~Cardoso, O.~J.~C. Dias, and L.~Gualtieri, ``The return of the membrane
  paradigm? {B}lack holes and strings in the water tap,''
\href{http://arXiv.org/abs/arXiv:0705.2777 [hep-th]}{{\tt arXiv:0705.2777
  [hep-th]}}.

\bibitem{obeHarmark:2005vn}
T.~Harmark and P.~Olesen, ``On the structure of stationary and axisymmetric
  metrics,'' {\em Phys. Rev.} {\bf D72} (2005) 124017,
\href{http://arXiv.org/abs/hep-th/0508208}{{\tt hep-th/0508208}}.

\bibitem{obeElvang:2003mj}
H.~Elvang and R.~Emparan, ``Black rings, supertubes, and a stringy resolution
  of black hole non-uniqueness,'' {\em JHEP} {\bf 11} (2003) 035,
\href{http://arXiv.org/abs/hep-th/0310008}{{\tt hep-th/0310008}}.

\bibitem{obeHovdebo:2006jy}
J.~L. Hovdebo and R.~C. Myers, ``Black rings, boosted strings and
  {Gregory-Laflamme},'' {\em Phys. Rev.} {\bf D73} (2006) 084013,
\href{http://arXiv.org/abs/hep-th/0601079}{{\tt hep-th/0601079}}.

\bibitem{obeElvang:2006dd}
H.~Elvang, R.~Emparan, and A.~Virmani, ``Dynamics and stability of black
  rings,'' {\em JHEP} {\bf 12} (2006) 074,
\href{http://arXiv.org/abs/hep-th/0608076}{{\tt hep-th/0608076}}.

\bibitem{obeCarter:2000wv}
B.~Carter, ``Essentials of classical brane dynamics,'' {\em Int. J. Theor.
  Phys.} {\bf 40} (2001) 2099--2130,
\href{http://arXiv.org/abs/gr-qc/0012036}{{\tt gr-qc/0012036}}.

\bibitem{obeKastor:2007wr}
D.~Kastor, S.~Ray, and J.~Traschen, ``The first law for boosted {Kaluza--Klein}
  black holes,'' {\em JHEP} {\bf 06} (2007) 026,
\href{http://arXiv.org/abs/arXiv:0704.0729 [hep-th]}{{\tt arXiv:0704.0729
  [hep-th]}}.

\bibitem{obeArcioni:2004ww}
G.~Arcioni and E.~Lozano-Tellechea, ``Stability and critical phenomena of black
  holes and black rings,'' {\em Phys. Rev.} {\bf D72} (2005) 104021,
\href{http://arXiv.org/abs/hep-th/0412118}{{\tt hep-th/0412118}}.

\bibitem{obeMaeda:2006hd}
K.-i. Maeda, N.~Ohta, and M.~Tanabe, ``A supersymmetric rotating black hole in
  a compactified spacetime,'' {\em Phys. Rev.} {\bf D74} (2006) 104002,
\href{http://arXiv.org/abs/hep-th/0607084}{{\tt hep-th/0607084}}.

\bibitem{obeKarlovini:2005cn}
M.~Karlovini and R.~von Unge, ``Charged black holes in compactified
  spacetimes,'' {\em Phys. Rev.} {\bf D72} (2005) 104013,
\href{http://arXiv.org/abs/gr-qc/0506073}{{\tt gr-qc/0506073}}.

\bibitem{obeCopsey:2006br}
K.~Copsey and G.~T. Horowitz, ``Gravity dual of gauge theory on {$S^2 \times
  S^1 \times \mathbb{R}$},'' {\em JHEP} {\bf 06} (2006) 021,
\href{http://arXiv.org/abs/hep-th/0602003}{{\tt hep-th/0602003}}.

\bibitem{obeMann:2006yi}
R.~B. Mann, E.~Radu, and C.~Stelea, ``Black string solutions with negative
  cosmological constant,'' {\em JHEP} {\bf 09} (2006) 073,
\href{http://arXiv.org/abs/hep-th/0604205}{{\tt hep-th/0604205}}.

\bibitem{obeKleihaus:2007dg}
B.~Kleihaus, J.~Kunz, and E.~Radu, ``Rotating nonuniform black string
  solutions,'' {\em JHEP} {\bf 05} (2007) 058,
\href{http://arXiv.org/abs/hep-th/0702053}{{\tt hep-th/0702053}}.

\bibitem{obeKleihaus:2007kc}
B.~Kleihaus, J.~Kunz, and F.~Navarro-Lerida, ``Rotating black holes in higher
  dimensions,''
\href{http://arXiv.org/abs/arXiv:0710.2291 [hep-th]}{{\tt arXiv:0710.2291
  [hep-th]}}.

\bibitem{obeKunduri:2006uh}
H.~K. Kunduri, J.~Lucietti, and H.~S. Reall, ``Do supersymmetric anti-de
  {Sitter} black rings exist?,'' {\em JHEP} {\bf 02} (2007) 026,
\href{http://arXiv.org/abs/hep-th/0611351}{{\tt hep-th/0611351}}.

\bibitem{obeElvang:2003yy}
H.~Elvang, ``A charged rotating black ring,''
\href{http://arXiv.org/abs/hep-th/0305247}{{\tt hep-th/0305247}}.

\bibitem{obeElvang:2004rt}
H.~Elvang, R.~Emparan, D.~Mateos, and H.~S. Reall, ``A supersymmetric black
  ring,'' {\em Phys. Rev. Lett.} {\bf 93} (2004) 211302,
\href{http://arXiv.org/abs/hep-th/0407065}{{\tt hep-th/0407065}}.

\bibitem{obeDabholkar:2006za}
A.~Dabholkar, N.~Iizuka, A.~Iqubal, A.~Sen, and M.~Shigemori, ``Spinning
  strings as small black rings,'' {\em JHEP} {\bf 04} (2007) 017,
\href{http://arXiv.org/abs/hep-th/0611166}{{\tt hep-th/0611166}}.

\bibitem{obeHelfgott:2005jn}
C.~Helfgott, Y.~Oz, and Y.~Yanay, ``On the topology of black hole event
  horizons in higher dimensions,'' {\em JHEP} {\bf 02} (2006) 025,
\href{http://arXiv.org/abs/hep-th/0509013}{{\tt hep-th/0509013}}.

\bibitem{obeGalloway:2005mf}
G.~J. Galloway and R.~Schoen, ``A generalization of {H}awking's black hole
  topology theorem to higher dimensions,'' {\em Commun. Math. Phys.} {\bf 266}
  (2006) 571--576,
\href{http://arXiv.org/abs/gr-qc/0509107}{{\tt gr-qc/0509107}}.

\bibitem{obeLahiri:2007ae}
S.~Lahiri and S.~Minwalla, ``Plasmarings as dual black rings,''
\href{http://arXiv.org/abs/arXiv:0705.3404 [hep-th]}{{\tt arXiv:0705.3404
  [hep-th]}}.

\bibitem{obeHarmark:2006df}
T.~Harmark, K.~R. Kristjansson, N.~A. Obers, and P.~B. Ronne, ``Three-charge
  black holes on a circle,'' {\em JHEP} {\bf 01} (2007) 023,
\href{http://arXiv.org/abs/hep-th/0606246}{{\tt hep-th/0606246}}.

\bibitem{obeHarmark:2007uy}
T.~Harmark, K.~R. Kristjansson, N.~A. Obers, and P.~B. Ronne, ``{Entropy of
  three-charge black holes on a circle},'' {\em Fortsch. Phys.} {\bf 55} (2007)
  748--753,
\href{http://arXiv.org/abs/hep-th/0701070}{{\tt hep-th/0701070}}.

\bibitem{obeChowdhury:2006qn}
B.~D. Chowdhury, S.~Giusto, and S.~D. Mathur, ``A microscopic model for the
  black hole - black string phase transition,'' {\em Nucl. Phys.} {\bf B762}
  (2007) 301--343,
\href{http://arXiv.org/abs/hep-th/0610069}{{\tt hep-th/0610069}}.

\bibitem{obeArkani-Hamed:1998rs}
N.~Arkani-Hamed, S.~Dimopoulos, and G.~R. Dvali, ``The hierarchy problem and
  new dimensions at a millimeter,'' {\em Phys. Lett.} {\bf B429} (1998)
  263--272,
\href{http://arXiv.org/abs/hep-ph/9803315}{{\tt hep-ph/9803315}}.

\bibitem{obeRandall:1999ee}
L.~Randall and R.~Sundrum, ``A large mass hierarchy from a small extra
  dimension,'' {\em Phys. Rev. Lett.} {\bf 83} (1999) 3370--3373,
\href{http://arXiv.org/abs/hep-ph/9905221}{{\tt hep-ph/9905221}}.

\bibitem{obeRandall:1999vf}
L.~Randall and R.~Sundrum, ``An alternative to compactification,'' {\em Phys.
  Rev. Lett.} {\bf 83} (1999) 4690--4693,
\href{http://arXiv.org/abs/hep-th/9906064}{{\tt hep-th/9906064}}.

\end{thebibliography}

\providecommand{\href}[2]{#2}\begingroup\raggedright\endgroup

\end{document}